\documentclass[a4paper,10pt]{article}
\usepackage[utf8]{inputenc}
\usepackage[T1]{fontenc}
\usepackage[a4paper,margin=1in,footskip=0.25in]{geometry}
\usepackage{graphicx}
\usepackage{tabularx}
\usepackage{array}
\usepackage{longtable}
\usepackage{multicol}
\usepackage{indentfirst,csquotes}
\usepackage{upgreek}
\usepackage{hhline}

\title{Generalized ladder operators and construction of families of superintegrable systems with Painlev\'e VI }

\author{Ian Marquette$^{1}$ and Anthony J. L. Parr$^{2}$}

\date{$^{1}$ Department of Mathematical and Physical Sciences, La Trobe University, Bendigo 3552, Victoria, Australia \\
$^{2}$ School of Mathematics and Physics, The University of Queensland, Brisbane 4072, Queensland, Australia \\
\texttt{i.marquette@latrobe.edu.au}; \texttt{anthony.parr@uq.edu.au}}
\usepackage{amsmath,amssymb,amsfonts,amsthm,appendix}
\usepackage{mathtools}

\usepackage{enumitem}
\newcommand{\mrm}[1]{\mathrm{#1}}

\newcommand\der[3]{\frac{\mrm{d}^{#1}#2}{\mrm{d}{#3}^{#1}}}
\usepackage{stmaryrd}
\theoremstyle{plain}

\newtheorem{theorem}{Theorem}[section]
\newtheorem{proposition}[theorem]{Proposition}

\newtheorem{lemma}[theorem]{Lemma}

\newtheorem*{corollary*}{Corollary}
\theoremstyle{definition}

\newtheorem*{definition*}{Definition}

\newtheorem*{example*}{Example}
\usepackage{paracol,bm}

\usepackage{mathrsfs}
\usepackage{microtype}
\usepackage{booktabs}
\usepackage[hidelinks]{hyperref}
\hypersetup{
  pdftitle={Generalized ladder operators and construction of families of superintegrable systems with Painle\'e VI},
  pdfauthor={Ian Marquette and Anthony J. L. Parr}
}
\usepackage[maxbibnames=99,backend=biber,sorting=none,giveninits=true,url=false,isbn=false]{biblatex}
\AtEveryBibitem{\clearfield{language}\clearfield{month}}

\numberwithin{equation}{section}
\allowdisplaybreaks[2]
\begin{document}
\maketitle
\begin{abstract}
We study quantum Hamiltonians whose separated angular component is expressed through a generic Painlev\'e VI transcendent.  The first part of the paper constructs generalized ladder operators for one-dimensional Schr\"odinger operators with a quadratic spectral variable. These operator algebras and their connection with Painlev\'e VI quantum Hamiltonians have not been explored in this setting. Providing these constructions is one of the main novelties of this paper. For the Painlev\'e VI angular Hamiltonian these operators satisfy $[N,J_\pm]=\pm2kJ_\pm$, where $N=\sqrt M$, and close through an eighth-degree norm polynomial.  Combining the angular shifts with the radial shift operators of the oscillator, Coulomb and free Bessel systems yields higher-order integrals and deformed Racah-type polynomial algebras for the corresponding two-dimensional superintegrable models. This construction provides families of Tremblay-Turbiner-Winternitz (TTW) and Post-Winternitz (PW) like Hamiltonians with integrals of arbitrarily high degree with Painlev\'e VI in the potential. Such Hamiltonians with integrals of arbitrary order and involving Painlev\'e VI are completely new. This framework also recovers models from the classification of third and fourth order integrals as particular cases. For a representative oscillator extension we compute the Casimir explicitly and recover the finite-dimensional energy spectrum from a deformed-oscillator realization. We then analyze the special-solution sector using the explicit families collected by Gromak, Laine and Shimomura.  The conventional Painlev\'e parameters are converted to the exponent variables of the quantum Hamiltonian, the resulting potentials are organized into a small number of spectral classes, and the distinction between formal zeros of the ladder norm and physical self-adjoint representations is made explicit. The regular nonclassical rational examples are rational Darboux deformations of the trigonometric P\"oschl--Teller system and lead to strictly isospectral exceptional-Jacobi sectors or to state-adding representations with an isolated angular level. The possibility of different types of levels, degeneracies and descriptions via representations offers the insight that models with separation in polar coordinates can display a much richer spectrum than the well-known hydrogen atom or radial oscillator, and can offer models of interest for applications in quantum and mathematical physics. A singular branch produces an interior inverse-square barrier and a split-domain problem.  For one regular exceptional sector and two state-adding sectors we also display the polynomial algebras explicitly.  These examples show how Painlev\'e VI special solutions control the representation structure and spectra of the associated polynomial symmetry algebras. 
\end{abstract}

\medskip
\noindent\textbf{Keywords:} Painlev\'e VI; quantum superintegrability; generalized ladder operators; polynomial algebras; exceptional Jacobi polynomials; Darboux transformations.

\smallskip
\noindent\textbf{MSC 2020:} 34M55; 81R12; 81Q05; 33E17.

\section{Introduction}

The Painlev\'e equations arose from the classification of second-order ordinary differential equations without movable critical singularities other than poles \cite{painleve1897,painleve02,gambier10,levi13,conte99,conte20}.  Painlev\'e VI occupies a distinguished position: it is the isomonodromic deformation equation of a rank-two Fuchsian system with four regular singular points, carries an affine-$D_4^{(1)}$ symmetry, and admits a rich monodromy and special-solution theory \cite{jim82,okamoto87pvi,iwasaki91,grom02}.  Its appearances range from integrable systems and random matrices to conformal field theory and geometric monodromy problems \cite{gamayun12, chenzhang10,boalch05,lisovyy14}.  The present work concerns a different but complementary role of $P_{\rm VI}$: the construction and spectral organization of quantum Hamiltonians whose separated potential depends on a Painlev\'e VI transcendent. Painlev\'e transcendents entered nonrelativistic quantum mechanics through the
study of higher-order intertwining, ladder and symmetry operators
\cite{fush,andrianov,carballo}.  The classification of two-dimensional systems
with higher-order integrals subsequently revealed potentials involving the first
three Painlev\'e transcendents without parameter restrictions \cite{gra04}, and
motivated the broader program of exotic quantum superintegrability
\cite{mar09,marquette11,mar10,mar17,mar19a,tre10,esc18a,esc17,esc18b,bern20b}.
All six Painlev\'e equations arise in this setting.  For reference, we use the
following conventional normalizations:
\begin{align}
 P_{\mathrm I}''(z)
 &=6P_{\mathrm I}(z)^2+z,
 \tag{$P_{\mathrm I}$}\\[1mm]
 P_{\mathrm{II}}''(z)
 &=2P_{\mathrm{II}}(z)^3+zP_{\mathrm{II}}(z)+\alpha,
 \tag{$P_{\mathrm{II}}$}\\[1mm]
 P_{\mathrm{III}}''(z)
 &=\frac{P_{\mathrm{III}}'(z)^2}{P_{\mathrm{III}}(z)}
   -\frac{P_{\mathrm{III}}'(z)}{z}
   +\frac{\alpha P_{\mathrm{III}}(z)^2+\beta}{z}
   +\gamma P_{\mathrm{III}}(z)^3
   +\frac{\delta}{P_{\mathrm{III}}(z)},
 \tag{$P_{\mathrm{III}}$}\\[1mm]
 P_{\mathrm{IV}}''(z)
 &=\frac{P_{\mathrm{IV}}'(z)^2}{2P_{\mathrm{IV}}(z)}
   +\frac32 P_{\mathrm{IV}}(z)^3
   +4zP_{\mathrm{IV}}(z)^2
   +2\bigl(z^2-\alpha\bigr)P_{\mathrm{IV}}(z)
   +\frac{\beta}{P_{\mathrm{IV}}(z)},
 \tag{$P_{\mathrm{IV}}$}\\[1mm]
 P_{\mathrm V}''(z)
 &=\left(\frac{1}{2P_{\mathrm V}(z)}
   +\frac{1}{P_{\mathrm V}(z)-1}\right)P_{\mathrm V}'(z)^2
   -\frac{P_{\mathrm V}'(z)}{z}\notag\\
 &\quad
   +\frac{\bigl(P_{\mathrm V}(z)-1\bigr)^2}{z^2}
    \left(\alpha P_{\mathrm V}(z)+\frac{\beta}{P_{\mathrm V}(z)}\right)
   +\frac{\gamma P_{\mathrm V}(z)}{z}
   +\frac{\delta P_{\mathrm V}(z)\bigl(P_{\mathrm V}(z)+1\bigr)}
    {P_{\mathrm V}(z)-1},
 \tag{$P_{\mathrm V}$}\\[1mm]
 P_{\mathrm{VI}}''(z)
 &=\frac12\left(
   \frac{1}{P_{\mathrm{VI}}(z)}
   +\frac{1}{P_{\mathrm{VI}}(z)-1}
   +\frac{1}{P_{\mathrm{VI}}(z)-z}\right)P_{\mathrm{VI}}'(z)^2\notag\\
 &\quad
   -\left(\frac1z+\frac{1}{z-1}
   +\frac{1}{P_{\mathrm{VI}}(z)-z}\right)P_{\mathrm{VI}}'(z)\notag\\
 &\quad
   +\frac{P_{\mathrm{VI}}(z)\bigl(P_{\mathrm{VI}}(z)-1\bigr)
   \bigl(P_{\mathrm{VI}}(z)-z\bigr)}{z^2(z-1)^2}
   \left[
   \alpha+\frac{\beta z}{P_{\mathrm{VI}}(z)^2}
   +\frac{\gamma (z-1)}{\bigl(P_{\mathrm{VI}}(z)-1\bigr)^2}
   +\frac{\delta z(z-1)}{\bigl(P_{\mathrm{VI}}(z)-z\bigr)^2}
   \right].
 \tag{$P_{\mathrm{VI}}$}
\end{align}
The parameters in each equation are arbitrary.  On distinguished parameter loci these
nonlinear equations admit rational, Riccati/hypergeometric, elliptic or algebraic
solutions. Away from such loci their transcendents are not reducible to solutions
of linear differential equations.

The term \emph{exotic potential} refers here to a Schr\"odinger potential that
contains a Painlev\'e transcendent rather than an elementary or classical special
function.  In contrast with the Lam\'e and hypergeometric settings, a generic
Painlev\'e-dependent spectral equation is not algebraized by a change of variable.
On a Riemannian manifold one may write schematically
\begin{equation}
 H\psi=E\psi,\qquad
 H=-\frac{\hbar^2}{2m}\frac{1}{\sqrt g}
 \partial_j\!\left(\sqrt g\,g^{jk}\partial_k\right)
 +V\bigl(x,P_J(x),P_J'(x)\bigr),
 \qquad J\in\{\mathrm I,\ldots,\mathrm{VI}\}.
\label{eq:schematic-exotic-Hamiltonian}
\end{equation}
For generic Painlev\'e parameters, asymptotic and connection information is then
needed to decide square integrability and self-adjoint boundary conditions.  On
special-solution loci, by contrast, the same Hamiltonian family may reduce to
rational or classical-function potentials.  This coexistence of generic and special
sectors is one reason that the spectral and representation patterns can vary sharply
within a single Painlev\'e family.

A complementary algebraic viewpoint starts from one-dimensional operator
relations.  Higher-order nonlinear equations and their Painlev\'e reductions enter
naturally in the construction of Abelian, Heisenberg, conformal and ladder algebras
\cite{Bureau64,Bureau64b,bureau71,cos01,cosgrove00,cosgrove91,mar19a}.  In the
ladder case, commensurable one-dimensional shifts can be combined to produce
superintegrable systems in two or more dimensions and integrals of arbitrarily high
order.  Rational fourth-Painlev\'e examples already show that changing the Painlev\'e
parameters can change the representation content and replace classical Hermite
polynomials by exceptional families \cite{hussin22,zel22}.

A Painlev\'e VI Hamiltonian was previously analyzed on particular rational loci for
which its wavefunctions are related to exceptional Jacobi polynomials
\cite{rita,ritb}.  The present paper has two linked aims.  First, we construct the
missing generalized ladder algebra for the one-dimensional $P_{\rm VI}$ angular
block and use the building-block method of Refs.~\cite{parr24,parr25} to obtain
higher-order integrals and deformed Racah-type algebras for oscillator, Coulomb and
free Bessel radial extensions.  Second, we determine what the explicit special solutions
collected by Gromak, Laine and Shimomura imply for the angular potential, the
self-adjoint spectrum and the representation theory.  The latter analysis is essential:
a zero of a formal norm polynomial is not automatically a physical lowest weight.

The resulting models illustrate how higher transcendental potentials can support
spectra and reducible modules that are absent from the elementary oscillator and
Coulomb systems, while retaining an exact polynomial-algebra description.

The direct derivation of higher-order superintegrable systems from their determining nonlinear partial differential equations rapidly becomes cumbersome, even after separation of variables. A complementary route is to classify one-dimensional operator algebras first and then combine compatible one-dimensional building blocks. This approach also exposes the algebraic structure and spectrum directly. To date, the most systematic classification of operator algebras is that of Marquette and Sajedi \cite{saj19}. They divided the candidate algebras into four types (up to scaling): Abelian (\([H,K]=0\)), Heisenberg (\([H,K]=1\)), conformal (\([H,K]=H\)) and ladder (\([H,K]=K\)). They considered operators \(K\) with orders up to and including five. At fifth order, additional nonlinear equations occur beyond the classical Painlev\'e list and their special-function content is less completely understood. From a physical perspective, Abelian, Heisenberg and conformal potential types give rise to continuous spectra on the whole real line or transcendental eigenvalues when the domain is bounded or semi-infinite. In contrast, the potentials of ladder type are well known to have quantised eigenvalues given by an arithmetic sequence. Under certain conditions, two Hamiltonians \(H_1,H_2\) admitting operator algebras of the aforementioned forms can be combined to form a two-dimensional superintegrable model \(H_1+H_2\) which separates in Cartesian coordinates. The results obtained for orders up to four are organised in the following array.
\begin{table}[ht]
\centering
\caption{Potentials possessing operator algebras in the classification of Ref.~\cite{saj19}.}
\label{tab:operator-algebra-classification}
\small
\begin{tabular}{@{}lcccc@{}}
\toprule
& \multicolumn{4}{c}{Order}\\
\cmidrule(lr){2-5}
Type & 1 & 2 & 3 & 4\\
\midrule
Abelian & \multicolumn{2}{c}{Free} & \multicolumn{2}{c}{Weierstrass}\\
Heisenberg & \multicolumn{2}{c}{Linear} & \multicolumn{2}{c}{Painlev\'e I}\\
Conformal & N/A & Inverse square & \multicolumn{2}{c}{Painlev\'e II}\\
Ladder & Square & Square \(+\) inverse square & Painlev\'e IV & Painlev\'e V\\
\bottomrule
\end{tabular}
\end{table}

Potentials involving Painlev\'e III and VI transcendents do not occur in that ordinary differential-ladder classification, but they do arise in separated two-dimensional superintegrability.  The key observation used here is that a separated Hamiltonian may have eigenvalues quadratic in the quantum number.  Its natural shift operators then depend on an adjoined square root of the Hamiltonian and are therefore formal rather than purely differential.  This enlarged operator setting recovers the Painlev\'{e} VI angular potential and leads, after suitable combinations, to actual higher-order integrals.

The paper is organized as follows.  Sections~2--6 develop the generalized one-dimensional ladders and their norm polynomials. The construction of such operators is one of the main results of this paper. Sections~7 and 8 construct the superintegrable extensions and discuss reduction of order. This second part of the paper, which builds on the ladder operators, is another of the main results, as it establishes that models involving Painlev\'e VI exist with integrals of arbitrarily high order and that TTW- and PW-like systems with Painlev\'e VI can be constructed, together with their integrals and polynomial algebras. Section~\ref{sec:PVI-special-solutions} organizes the explicit Gromak--Laine--Shimomura special solutions into potential and spectral classes, while Section~\ref{sec:exceptional-Jacobi-Gromak} identifies the exceptional-Jacobi/Darboux sectors and their reducible ladder representations.  Section~\ref{sec:conclusion} summarizes the principal consequences and open problems.

\section{Formal Ladder Operators of Hamiltonians with Quadratic Eigenvalues}
\label{sec:formal-ladders}
This section sets up the algebraic framework used throughout the paper. When the eigenvalues of $M$ are quadratic in the quantum number, consecutive levels are not equally spaced: the gap between them, $4k(N+k)$, depends on the square root $N=\sqrt M$. A ladder relation with a constant step therefore holds for $N$ rather than for $M$, and the corresponding ladder operators depend on $N$. We fix the branch of $N$, distinguish formal operators from actual ones, derive the equations that a ladder operator must satisfy, and introduce the norm polynomials $\Phi^{(n)}$, whose roots are the candidate lowest and highest weights.

We consider Schr\"{o}dinger operators
\[M=-\partial_\theta^2+V(\theta)\]
whose eigenvalues form an arithmetic sequence of degree 2,
\[M\Theta_m(\theta)=(2km+\lambda)^2\Theta_m(\theta),\qquad m=0,1,2,\ldots.\]
We adjoin a square-root \(N\coloneqq\sqrt{M}\) by fixing the branch
\[N\Theta_m(\theta)=(2km+\lambda)\Theta_m(\theta).\]
This is an algebraic ansatz for an eigenfunction sequence, not an assertion that every such sequence is the spectrum of a self-adjoint realization. An operator will be called \textit{actual} if it can be resolved into a polynomial function of \(M\) and potentially other differential operators. We call an operator \textit{formal} if it is a rational function of \(N\) or, if properly polynomial, not invariant under the involution \(N\mapsto-N\). We start the construction by using ladder operators \(J_\pm\) that satisfy
\begin{equation}
\left[N,J_{\pm}\right]=\pm 2kJ_{\pm}\label{eq:lad1}
\end{equation}
and by splitting each of them into two blocks, writing
\[J_\pm=J_a\pm J_fN.\]
We then obtain the equivalent coupled system
\begin{subequations}
    \begin{align}
[M,J_a]&=4k(kJ_a+J_fM),\\
[M,J_f]&=4k(J_a+kJ_f).
    \end{align}
\end{subequations}
and therefore
\begin{equation}
[M,J_\pm]=4kJ_\pm\left(k\pm N\right)
\end{equation}
We may take \(J_a,J_f\) to be invariant under sign change of \(N\) and hence \(J_-=J_+|_{N\mapsto -N}\). Below we often write \(J=J_+\). The relations above will prove to be quite useful in connecting Painlev\'e VI with an underlying operator algebra in the context of quantum mechanics. They also demonstrate that, unlike the previous cases defined by Abelian, Heisenberg, conformal and ladder algebras, other types of structures are of interest for connecting transcendental functions defined via nonlinear ODEs to spectral problems in quantum mechanics. We restrict attention to ladders for which \(J_+J_-\) is a scalar polynomial in \(N\). The term norm polynomial refers to this algebraic product. It does not assume \(J_-=J_+^\dagger\). A positive Hilbert-space norm requires a compatible adjoint convention and operator domain. Define the norm polynomials

\[\Phi^{(n)}\left(N\right)\coloneqq J^n_+J^n_-,\qquad \Phi^{(0)}\coloneqq 1,\qquad \Phi=\Phi^{(1)}.\]
They obey
\[\Phi^{(n+1)}\left(N\right)=\Phi^{(n)}\left(N-2k\right)\Phi\left(N\right),\qquad  J^n_-J^n_+=\Phi^{(n)}\left(-N\right)=\Phi^{(n)}\left(N+2kn\right)\]

The roots of \(\Phi^{(1)}\) are algebraic candidates for the lowest- or highest-weight values of \(N\). They correspond to physical states, such as ground states of \(M\), only after a domain, self-adjoint boundary conditions, normalizability and positivity of the norm have been imposed. One further identity will be needed. Evaluating \(J^n\) at \(N=-nk\) and reducing modulo the eigenvalue equation \(M=n^2k^2\) gives
\begin{equation}
\Phi^{(n)}(nk)=\left(\left.J^n\right|_{N\mapsto -nk}\right)^2\mod\left(n^2k^2+\partial_\theta^2-V\right).
\end{equation}

When \(J^n|_{N\mapsto -nk}\) is a constant, this fixes it up to sign. These special values control when \(J^n\) can be reduced in order (Theorem~\ref{th:reduce}). These relations are formal within the sign-reflected, scalar norm-polynomial setting specified above. They do not by themselves establish a physical ladder representation. In the next section we write \(J\) as a polynomial in \(N\) and \(\partial_\theta\) and turn \eqref{eq:lad1} into a recursive system for its coefficients. The following sections then determine, order by order, which potentials \(V\) admit such ladder operators.

\subsection{Darboux transformations and ladder operators}
\label{sec:darboux}
The examples of order reduction below can be understood using Darboux
transformations.  We recall the relevant factorization, in the notation of
Section~\ref{sec:formal-ladders} \cite{darboux82,crum55,que15,ull14}. For the real-potential case, suppose that \(\psi_0\) is a real, nonvanishing solution of
\(M\psi_0=\varepsilon\psi_0\) in the interior of the interval, where \(\varepsilon=\nu_0^2\), and set \(w=\psi_0'/\psi_0\).  The first-order operators
\begin{equation}
 A=\partial_\theta-w,\qquad A^\dagger=-\partial_\theta-w
 \label{eq:darboux-intertwiners}
\end{equation}
satisfy the following factorization relations
\begin{equation}
 \begin{aligned}
 A^\dagger A&=M-\varepsilon, &
 AA^\dagger&=\widetilde M-\varepsilon,\\
 \widetilde M&=-\partial_\theta^2+\widetilde V, &
 \widetilde V&=V-2(\log\psi_0)'',
 \end{aligned}
 \label{eq:darboux-factorizations}
\end{equation}
and consequently the intertwining relations
\begin{equation}
 \widetilde M A=AM,\qquad MA^\dagger=A^\dagger\widetilde M.
 \label{eq:darboux-intertwining}
\end{equation}
Here the dagger denotes the formal adjoint. It denotes an adjoint of self-adjoint realizations only when the corresponding domains are chosen appropriately. The solutions \(\psi_0\) and \(\psi_0^{-1}\) lie in the kernels of \(A\) and \(A^\dagger\), respectively. A physical energy level can be deleted or added only when the relevant solution belongs to the domain of the corresponding self-adjoint operator. In particular, square integrability alone does not guarantee that the required boundary conditions are satisfied. For a chain with \(r\) seeds \(\psi_1,\ldots,\psi_r\), the Crum operator \(\mathcal B_r\) and its transformed potential satisfy
\begin{equation}
 \widetilde V=V-2\frac{d^2}{d\theta^2}
       \log W(\psi_1,\ldots,\psi_r),\qquad
 \mathcal B_r^\dagger\mathcal B_r
       =\prod_{i=1}^r(M-\varepsilon_i).
 \label{eq:darboux-crum}
\end{equation}

The displayed Wronskian formula requires the Wronskian to have no zeros in the interior. An intertwiner also transports ladder operators to the transformed Hamiltonian. We write \(\widetilde N=\sqrt{\widetilde M}\), using the same positive spectral convention as for \(N=\sqrt M\) on the corresponding eigenstates. On the span of transported eigenfunctions, the positive square roots satisfy \(\widetilde N A=AN\). This does not by itself identify the self-adjoint domains. If \([N,J_\pm]=\pm 2kJ_\pm\) and
\(J_+J_-=\Phi(N)\), define
\begin{equation}
 \widetilde J_\pm=AJ_\pm A^\dagger.
 \label{eq:darboux-dressed-ladders}
\end{equation}
On the common spectral domain these have the same formal step
\([\widetilde N,\widetilde J_\pm]=\pm2k\widetilde J_\pm\), and
\begin{equation}
 \begin{aligned}
 \widetilde J_+\widetilde J_-
 &=\Phi(\widetilde N)
      (\widetilde N^2-\nu_0^2)
      [ (\widetilde N-2k)^2-\nu_0^2]\\
 &=\Phi(\widetilde N)
   [(k-\nu_0)^2-(\widetilde N-k)^2]
   [(k+\nu_0)^2-(\widetilde N-k)^2].
 \end{aligned}
 \label{eq:darboux-dressed-norm}
\end{equation}
Indeed, insert \(A^\dagger A=M-\varepsilon\) between
\(J_+\) and \(J_-\), use
\((M-\varepsilon)J_-=J_-[(N-2k)^2-\varepsilon]\),
and then use \eqref{eq:darboux-intertwining} and
\(AA^\dagger=\widetilde N^2-\varepsilon\).
Thus a Darboux transformation preserves the shifted norm-polynomial form \eqref{eq:normpol}, adding the two formal parameters \(1\pm\nu_0/k\).  The dressed expression has formal order two higher (counting \(N\) as first order) before possible right factors in the spectral variable are removed. These factors, and the associated reduction of order, will be made
explicit in Section~\ref{subsec:darboux-chain-reduction}.  The relations
are formal at spectral thresholds where the positive square-root branch or the selected operator domain is not preserved. For a chain with no seeds, \(r=0\), the empty Wronskian and product equal one. The relation \eqref{eq:darboux-dressed-norm} gives the algebraic product of the dressed ladders.

\section{Construction of Ladder Operators}
\label{sec:construction}
We now use further insight into the structure of the operator $J$, and an expansion in terms of \(N\) and $\partial_\theta$, to make the ladder relation \eqref{eq:lad1} explicit. We write \(J\) as a polynomial in \(N\) whose coefficients are functions of \(\theta\) and first-order differential operators. Then \eqref{eq:lad1} becomes a recursive system of linear differential equations for these coefficients, coupled to the potential \(V\). We solve the first two steps of this recursion in general and show that they fix the leading terms of \(J\) up to a single constant. We then record the form of the norm polynomial found in all the cases treated below. Finally, we prove a criterion (Theorem~\ref{th:reduce}) for powers of \(J\) to admit reductions of order. The order \(\mu\) of \(J\) is fixed in Sections~4--6.

Let us expand the ladder operator as a polynomial in \(N\):
\begin{equation}
J=\sum^\mu_{i=0}F_i(\theta)N^{\mu-i}+\sum^{\mu-1}_{i=0}G_i(\theta)\partial_\theta N^{\mu-1-i}
\end{equation}
where the coefficients \(F_i,G_i\) are to be determined and \(\mu\) is the order of \(J\). Then 
\begin{equation}
    [M,J]=4kJ\left(N+k\right)\label{eq:recequ}
\end{equation}
transforms into
\begin{subequations}
\begin{align}
    4kF_j+4k^2F_{j-1}&=-\der{2}{F_{j-1}}{\theta}+2\der{}{G_j}{\theta}-2V\der{}{G_{j-2}}{\theta}-G_{j-2}\der{}{V}{\theta};\label{eq:system1}\\
    4kG_j+4k^2G_{j-1}&=-\der{2}{G_{j-1}}{\theta}-2\der{}{F_{j}}{\theta};\label{eq:system2}\\
    F_j=0,&\qquad j<0,\, j>\mu;\\
    G_j=0,&\qquad j<0,\, j\geq\mu.
\end{align}
\end{subequations}
In particular, when \(j=0\), \eqref{eq:system1}--\eqref{eq:system2} reads
\begin{align*}
4kF_0&=2\der{}{G_0}{\theta},\\
4kG_0&=-2\der{}{F_0}{\theta}.
\end{align*}
There are two degrees of freedom in \eqref{eq:recequ}, namely scaling of \(J\) and translation of \(\theta\). These are fixed upon the choice of solution for \(F_0,G_0\). We take them to be
\begin{equation}
F_0=\cos(2k\theta),\qquad G_0=\sin(2k\theta).\label{eq:a0}
\end{equation}
When \(j=1\), we have from \eqref{eq:system1}--\eqref{eq:system2}:
\begin{align*}
    4kF_1&=2\der{}{G_1}{\theta}\\
    4kG_1&=-2\der{}{F_1}{\theta}
\end{align*}
from which we must use the general solution
\begin{equation}
    F_1=c_1\cos(2k\theta)+c_2\sin(2k\theta),\qquad G_1=c_1\sin(2k\theta)-c_2\cos(2k\theta)\label{eq:a1}
\end{equation}
where \(c_1,c_2\) are constants. 
\begin{proposition}
The constant \(c_2\) is equal to zero.
\end{proposition}
\begin{proof}
Let \(t\) be an indeterminate that commutes with \(\theta,\partial_\theta\). If we evaluate
\begin{align*}
    \Phi(t)&=J_+J_-|_{N\mapsto t}=J|_{N\mapsto t-2k}J
    |_{N\mapsto -t}\mod(t^2+\partial_\theta^2-V)\\
    &=(-1)^\mu t^{2\mu-4}\{t[\cos(2k\theta)(t-2\mu k+c_1)+c_2\sin(2k\theta)]\\
    &\qquad+[\sin(2k\theta)(t-2\mu k+2k+c_1)-c_2\cos(2k\theta)]\partial_\theta\}\\
    &\qquad\times\{t[\cos(2k\theta)(t-c_1)-c_2\sin(2k\theta)]\\
    &\qquad-[\sin(2k\theta)(t-c_1)+c_2\cos(2k\theta)]\partial_\theta\}+\mathcal{O}(t^{2\mu-2},t^{2\mu-3}\partial_\theta)\mod(t^2+\partial_\theta^2-V)\\
    &=(-1)^\mu (t^{2\mu}-2\mu k t^{2\mu-1}-2c_2 t^{2\mu-2}\partial_\theta)+\mathcal{O}(t^{2\mu-2},t^{2\mu-3}\partial_\theta)
\end{align*}
Then the coefficient of \(t^{2\mu-2}\partial_\theta\) must vanish, because \(\Phi\) is assumed to be a scalar polynomial in \(N\). Hence \(c_2=0\). 
\end{proof}
For the ladder operators we examine below, we find the norm polynomial to be of the form
\begin{equation}
\Phi\left(N\right)=\prod^\mu_{j=1}\left[k^2\rho_j^2-\left(N-k\right)^2\right].\label{eq:normpol}
\end{equation}
where \(\rho_1,\rho_2,\ldots,\rho_\mu\) are constant parameters. Supposing this, we readily deduce
\begin{equation}
\Phi^{(n)}\left(N\right)=\prod^n_{i=1}\prod^\mu_{j=1}\left[k^2\rho_j^2-\left(N-2ki+k\right)^2\right].\label{eq:normpolext}
\end{equation}
\begin{lemma}
    Let 
     \begin{equation}
     \phi^{(n)}\coloneqq J^n|_{N\mapsto -nk}.
 \end{equation}
Products are first reduced using \(M=N^2\) to polynomials in rightmost \(N\), with differential coefficients of order at most one. After setting \(N=t\), write the resulting operator as \(a_n(\theta,t)\partial_\theta+b_n(\theta,t)\). Thus the evaluation is made as the first-order right remainder modulo \(M-t^2\). If \(\phi^{(n)}\) is a constant for a given \(n\), then \(\phi^{(n+2i)}\) is a constant for all \(i\in\mathbb{N}\). Moreover, if \(\Phi\) is of the form \eqref{eq:normpol} and \(\phi^{(n)}\) is a constant, then
 \begin{equation}
\phi^{(n)}=\pm k^{\mu n}\prod^n_{i=1}\prod^\mu_{j=1}\left(\rho_j-n-1+2i\right)\label{eq:phiform}
\end{equation}
\end{lemma}
\begin{proof}
Suppose \(\phi^{(n)}\) is a constant for some \(n\). We have the recurrence relation
\begin{align}
\phi^{(n+2)}&=JJ^nJ|_{N\mapsto -k(n+2)}\nonumber\\
&=J|_{N\mapsto kn}J^n|_{N\mapsto -kn}J|_{N\mapsto -k(n+2)}\mod(k^2(n+2)^2+\partial_\theta^2-V)\nonumber\\
&=\phi^{(n)}J_+J_-|_{N\mapsto k(n+2)}\nonumber\\
&=\phi^{(n)}\Phi(k(n+2)).
\end{align}
So \(\phi^{(n+2)}\) is also constant. This resolves the first part of the lemma. For the second, we have
\[\Phi^{(n)}(nk)=J^n_+|_{N\mapsto -nk}J^n_-|_{N\mapsto nk}\mod(n^2k^2+\partial_\theta^2-V)=\left[\phi^{(n)}\right]^2\]
Now,
\begin{align*}
\Phi^{(n)}\left(nk\right)&=k^{2\mu n}\prod^n_{i=1}\prod^\mu_{j=1}\left[\rho_j^2-(n-2i+1)^2\right]\\
&=k^{2\mu n}\prod^n_{i=1}\prod^\mu_{j=1}(\rho_j-n+2i-1)(\rho_j+n-2i+1)\\
&=k^{2\mu n}\prod^n_{i=1}\prod^\mu_{j=1}(\rho_j-n+2i-1)^2
\end{align*}
hence formula \eqref{eq:phiform}.
\end{proof}
We may modify the sign of \(J\) and/or the roots \(\rho_1,\rho_2,\ldots,\rho_\mu\) so that the sign in \eqref{eq:phiform} is always positive.

Let \(\Phi\) be of the form \eqref{eq:normpol}, with \(k\neq0\), and let \(\left(\phi^{(*+1)}\right)^{-1}(\mathbb{R})\) be the set of nonnegative integers \(u\) such that \(\phi^{(u+1)}\) is constant. Let
\begin{equation}
    \mathcal{P}\coloneqq\{\rho_j,-\rho_j\}^\mu_{j=1}\cap \left(\phi^{(*+1)}\right)^{-1}(\mathbb{R})
\end{equation}
We distinguish three cases. First, if \(\mathcal{P}\) is empty, we set \(S\coloneqq\varnothing\). Second, if \(\mathcal{P}\) contains integers of one parity only, then we set \(S\coloneqq\{\min \mathcal{P}\}\). Otherwise, set \(S\coloneqq\{s_e,s_o\}\) where \(s_e=\min(\mathcal P\cap2\mathbb Z)\), \(s_o=\min\bigl(\mathcal P\cap (2\mathbb Z+1)\bigr)\). Empty products below equal one.

\begin{theorem}
Let \(\Phi\) be of the form \eqref{eq:normpol} and \(S\) defined as above. Define
\begin{equation}
    D^{(n)}_S\left(N\right)\coloneqq \prod_{s\in S}\prod^{n-s-1}_{i=0}\left[N+k(s+2i+1)\right]
\end{equation}
Then the reduced ladder operators
\[L^{(n)}_\pm\coloneqq J^n_\pm\left[D^{(n)}_S\left(\pm N\right)\right]^{-1}\]
are polynomial in \(N\).\label{th:reduce}
\end{theorem}
\begin{proof}
Let \(s\in S\). Then \(\phi^{(s+2i+1)}=0\) for all \(i\geq 0\). Suppose \(n>s\). For \(0\leq i\leq\frac{1}{2}(n-s-1)\), we have
\begin{align*}
J^n|_{N\mapsto-k(s+2i+1)}&=J^{n-s-2i-1}J^{s+2i+1}|_{N\mapsto-k(s+2i+1)}\\
&=\phi^{(s+2i+1)}J^{n-s-2i-1}|_{N\mapsto k(s+2i+1)}\\
&=0.
\end{align*}
Similarly, for \(\frac{1}{2}(n-s-1)<i\leq n-s-1\)
\begin{align*}
J^n&|_{N\mapsto-k(s+2i+1)}=J^{2n-2i-s-1}J^{s+2i+1-n}|_{N\mapsto-k(s+2i+1)}
\\
&=J^{2n-2i-s-1}|_{N\mapsto -k(2n-s-2i-1)}J^{s+2i+1-n}|_{N\mapsto-k(s+2i+1)}\mod ( k^2 (2n-s-2i-1)^2 +\partial_\theta^2-V)\\
&=\phi^{(s+2(n-i-s-1)+1)}J^{s+2i+1-n}|_{N\mapsto-k(s+2i+1)}=0
\end{align*}
For each \(t=-k(s+2i+1)\), these compositions vanish on both independent local solutions of \(Mf=t^2f\). In the first-order normal form this forces \(a_n(\theta,t)=b_n(\theta,t)=0\), since the two solutions have nonzero Wronskian on a regular local interval. The scalar factor theorem therefore gives a right factor \(N-t\) in both coefficient polynomials. The roots are distinct because \(k\neq0\) and at most one index of each parity is used. Their product divides both coefficients, proving the stated polynomiality. The same argument applies after \(N\mapsto-N\).
\end{proof}

The reduced ladder operators satisfy the product rules
\begin{align}
L^{(m)}_\pm L^{(n)}_\pm&=L^{(m+n)}_\pm\,\frac{D^{(m+n)}_S(\pm N)}{D^{(m)}_S(\pm N+2kn)\,D^{(n)}_S(\pm N)},\\
L^{(n)}_+L^{(n)}_-&=\frac{\Phi^{(n)}(N)}{D^{(n)}_S(N-2kn)\,D^{(n)}_S(-N)},\\
L^{(n)}_-L^{(n)}_+&=\frac{\Phi^{(n)}(N+2kn)}{D^{(n)}_S(N)\,D^{(n)}_S(-N-2kn)},
\end{align}
which follow from \(f(N)J^n_\pm=J^n_\pm f(N\pm2kn)\) and the definition of \(\Phi^{(n)}\). Since \(L^{(n)}_\pm\) are polynomial in \(N\), the right-hand sides are polynomials. In particular, the norm of the reduced operators is \(\Phi^{(n)}\) with the factors \(D^{(n)}_S\) removed.

The remaining coefficients of \(J\) depend on its order \(\mu\). In the next three sections we solve the recursion for \(\mu=1\), \(\mu=2\) and \(\mu=3,4\). The first two cases give the P\"oschl--Teller family, and the third- and fourth-order cases lead to Painlev\'e VI.

\section{First-Order Ladder Operators}\label{firstorder}
We begin with the simplest case, a ladder operator of first order. We show that it forces the potential to be the symmetric P\"oschl--Teller potential, that \(J\) is then the familiar \(\mathfrak{sl}(2)\) shift operator, and that its norm polynomial has the form \eqref{eq:normpol} with a single parameter. This case is a check of the construction, and it supplies the angular block of the explicit quadratic algebra in Section~\ref{subsec:explicit-quadratic-algebra}.

Take \(\mu=1\), so that \(c_1=0\) and

\[J=\sin(2k\theta)\partial_\theta+\cos(2k\theta)N\]
For \eqref{eq:system1} when \(j=2\), we have
\[0=\sin(2k\theta)V'+4k\cos(2k\theta)V\]
from which we deduce
\begin{equation}
V(\theta)=\frac{k^2(\rho^2-1)}{\sin^2(2k\theta)}\label{eq:V1}
\end{equation}
The norm polynomial is easily determined to be
\[\Phi\left(N\right)=k^2\rho^2-\left(N-k\right)^2\]
which is of the form \eqref{eq:normpol}. The eigenfunctions are well-known
\begin{equation}
    \Theta_{n}(\theta)=[\sin(2k\theta)]^{\frac{\rho+1}{2}}\mathscr{C}_n^{\left(\frac{\rho+1}{2}\right)}\left(\cos(2k\theta)\right)
\end{equation}
where \(\mathscr{C}_n^{(\alpha)}\) are the Gegenbauer polynomials:
\begin{equation}
    \mathscr{C}^{(\alpha)}_n(x)=\frac{(-1)^n\Gamma\left(\alpha+\frac{1}{2}\right)\Gamma(n+2\alpha)}{2^nn!\Gamma(2\alpha)\Gamma\left(\alpha+n+\frac{1}{2}\right)}\left(1-x^2\right)^{\frac{1}{2}-\alpha}\frac{\mathrm{d}^n}{\mathrm{d}x^n}\left[\left(1-x^2\right)^{n+\alpha-\frac{1}{2}}\right]
\end{equation}


The eigenfunctions \(\Theta_{n}\) are square-integrable on \(\theta\in\left(0,\frac{\pi}{2k}\right)\), where \(\sin(2k\theta)>0\), provided \(\rho>-2\). For \(\rho>-1\) the branch of \(N\) fixed in Section~\ref{sec:formal-ladders} is the positive square root of \(M\). The ladder operators on this basis generate an \(\mathfrak{sl}(2)\)-module:
\begin{subequations}
\begin{align}
    M\Theta_{n}&=k^2(2n+1+ \rho)^2\Theta_{n}\\
    J_+\Theta_{n}&=2k(n+1)\Theta_{n+1}\\
    J_-\Theta_{n}&=\begin{cases}-2k(n+\rho)\Theta_{n-1},&\quad n>0;\\
    0,&\quad n=0,\end{cases}
    \end{align}
\end{subequations}
with
\[J_+J_-+J_-J_++2M=2k^2(\rho^2-1).\]
Thus the first-order ladder reproduces the familiar \(\mathfrak{sl}(2)\) structure of the symmetric P\"oschl--Teller potential.

We calculate some powers of the ladder operator
\begin{align*}
    J^2
    &=[\sin(4k\theta)\partial_\theta+\cos(4k\theta)N](N+2k)+k^2(\rho^2-1)\\
    J^3&=[\sin(6k\theta)\partial_\theta+\cos(6k\theta)N](N+2k)(N+4k)\\
    &\qquad+k^2(\rho^2-1)[\sin(2k\theta)\partial_\theta+\cos(2k\theta)(3N+8k)]\\
    J^4&=\{[\sin(8k\theta)\partial_\theta+\cos(8k\theta)N](N+2k)(N+6k)\\
    &\qquad+2k^2(\rho^2-1)[\sin(4k\theta)\partial_\theta+2\cos(4k\theta)(N+3k)\\
    &\qquad+N+4k]\}(N+4k)+k^4(\rho^2-1)(\rho^2-9)\\
J^5&=\{[\sin(10k\theta)\partial_\theta+\cos(10k\theta)N](N+2k)(N+8k)   \\
&\qquad+k^2(\rho^2-1)[3\sin(6k\theta)\partial_\theta+\cos(6k\theta)(5N+16k)]\}(N+4k)(N+6k)\\
&\qquad+k^2(\rho^2-1)[\sin(2k\theta)\partial_\theta+\cos(2k\theta)(5N+24k)][(N+5k)^2+k^2(\rho^2-10)]
\end{align*}
In particular, \(\phi^{(2)}=k^2(\rho^2-1)\) is constant and hence
 \[\phi^{(2n)}=k^{2n}\prod^n_{i=1}\left[\rho^2-(2i-1)^2\right]\]
 however \(\phi^{(2n+1)}\) is never constant for any \(n\). 
\subsection{Reduction of Order}
If \(\rho\) is an odd integer, a vanishing even-power remainder gives removable right factors in higher powers of \(J\), including odd powers. For \(\rho=\pm1\), the potential is zero. For \(\rho=\pm3\), the reductions
\[
 J^4=L^{(4)}(N+4k),\qquad
 J^5=L^{(5)}(N+4k)(N+6k)
\]
make the reduction explicit. The spectral factors act on the right, and the reduced operators are
 \begin{equation}
 \begin{aligned}L^{(4)}&=[\sin(8k\theta)\partial_\theta+\cos(8k\theta)N](N+2k)(N+6k)\\
    &\qquad+16k^2[\sin(4k\theta)\partial_\theta+2\cos(4k\theta)(N+3k)+N+4k]
\end{aligned}\label{eq:l4}
\end{equation}
\begin{equation}
\begin{aligned}
    L^{(5)}&=[\sin(10k\theta)\partial_\theta+\cos(10k\theta)N](N+2k)(N+8k)   \\
&\qquad+8k^2[3\sin(6k\theta)\partial_\theta+\cos(6k\theta)(5N+16k)]\\
&\qquad+8k^2[\sin(2k\theta)\partial_\theta+\cos(2k\theta)(5N+24k)]
    \end{aligned}\label{eq:l5}
    \end{equation}
Both displayed operators have formal order three, counting \(N\) as first order. For \(\rho=\pm3\), this order remains three for all \(n\geq4\). The Darboux construction explains why the order no longer grows with the step. We write
    \[M_m=-\partial_\theta^2+\frac{4k^2m(m-1)}{\sin^2(2k\theta)},\qquad \mathcal{A}_m=\partial_\theta-2km\cot(2k\theta)\]
    Then
    \begin{align*}
    \mathcal{A}_mM_m&=M_{m+1}\mathcal{A}_m\\
        \mathcal{A}_m^\dagger\mathcal{A}_m&=M_m-4k^2m^2\\
    \mathcal{A}_m\mathcal{A}_m^\dagger&=M_{m+1}-4k^2m^2
    \end{align*}
Thus \(M_1=-\partial_\theta^2\) is the free Hamiltonian, and \(\mathcal A_m\) maps from \(M_m\) to \(M_{m+1}\). A chain of \(m-1\) Darboux steps adds \(2(m-1)\) to the formal order of a dressed free ladder without changing its step. Since the triple \(J_a=\sin(2\alpha k\theta)\partial_\theta\), \(J_f=\cos(2\alpha k\theta)\), \(M_1=-\partial_\theta^2\) satisfies 
    \[[M_1,J_a]=4\alpha k(\alpha kJ_a+J_fM_1),\qquad [M_1,J_f]=4\alpha k(J_a+\alpha kJ_f)\]
    we have that
    \[L^{(\alpha)}_m=\left(\prod^{m-1}_{i=1}\mathcal{A}_{m-i}\right)\left[\sin(2\alpha k\theta)\partial_\theta+\cos(2\alpha k\theta)N_1\right]\left(\prod^{m-1}_{i=1}\mathcal{A}_i^\dagger\right)\]    
    is of formal order \(2m-1\) and is a ladder operator for \(M_m\), i.e.
    \[[N_m,L^{(\alpha)}_m]=2\alpha kL_m^{(\alpha)}\]
Here \(N_m=\sqrt{M_m}\), and the rightmost factor acts first. The relation is understood on the common transported eigenfunction span. On the free Dirichlet basis the step index \(\alpha\) is a positive integer. Expanding the dressed product for \(m=2\), so that \(\rho=3\), with all factors \(N_2\) on the right gives

\begin{equation}
\begin{aligned}
        L^{(\alpha)}_2&=[\sin(2\alpha k\theta)\partial_\theta+\cos(2\alpha k\theta)N_2](N_2+2k)[N_2+2(\alpha-1)k]\\&\qquad-2k^2\csc^2(2k\theta)\{(\alpha-2)\sin(2\alpha k\theta)-\alpha\sin[2(\alpha-2)k\theta]\}\partial_\theta\\
        &\qquad-k^2\csc^3(2k\theta)\{\sin[2k(\alpha+1)\theta][\alpha(N_2+4k)-4k]\\
        &\qquad-2\sin[2k(\alpha-1)\theta][\alpha(N_2+2k)+2k]+\alpha\sin[2k(\alpha-3)\theta]N_2\}
\end{aligned}
\end{equation}
For \(\alpha=n\in\mathbb N\), finite-angle identities turn the apparent trigonometric denominators into finite sums:
\begin{equation}
\begin{aligned}
        L^{(n)}_2&=[\sin(2n k\theta)\partial_\theta+\cos(2n k\theta)N_2](N_2+2k)[N_2+2(n-1)k]\\&\qquad+8k^2\sum^{\left\lfloor\frac{n}{2}\right\rfloor}_{i=1}(n-2i)\sin[2(n-2i)k\theta]\partial_\theta\\
        &\qquad+8k^2\sum^{\left\lfloor\frac{n}{2}\right\rfloor}_{i=1}\cos[2(n-2i)k\theta][4ki(n-i)+n N_2]\\
        &\qquad-2n k^2[1+(-1)^n](N_2+nk).
    \end{aligned}
\end{equation}
For \(n=4\) and \(n=5\), this coincides with \eqref{eq:l4} and \eqref{eq:l5}, respectively.

\section{Second-Order Ladder Operators}
\label{sec:second-order}
A second-order ladder operator allows the two singular terms of the potential to carry independent strengths. We show that the recursion then gives the general P\"oschl--Teller potential, whose eigenfunctions are built from Jacobi polynomials, and that \(J\) has a two-factor norm polynomial and satisfies the reduction criterion of Theorem~\ref{th:reduce}. When the two strengths coincide, \(J\) factorizes and reduces to the first-order operator of Section~\ref{firstorder}. The following subsections explain how integer norm parameters give one or two families of removable spectral factors, and how Darboux chains realize the corresponding reductions.

We now take \(\mu=2\). From \eqref{eq:system2} for \(j=2\), we have

\[4k^2G_1=-\der{2}{G_1}{\theta}-2\der{}{F_2}{\theta}\]
so \(F_2(\theta)=c_3\) is a constant. Substituting into \eqref{eq:system1}, we obtain
\[4kc_3+4k^2F_1=-\der{2}{F_1}{\theta}-4kV\cos(2k\theta)-\der{}{V}{\theta}\sin(2k\theta)\]
Solving for \(V\), we find the P\"{o}schl-Teller potential in its general form
\begin{align*}
    V&=\frac{c_4+2c_3\cos(2k\theta)}{\sin^2(2k\theta)}
\end{align*}
for some other constant \(c_4\).  To express this in a convenient way, we take \(c_3=k^2\rho_1\rho_2\) and \(c_4=k^2(\rho_1^2+\rho_2^2-1)\)
to give
\begin{equation}
    V(\theta)=\frac{k^2}{4}\left[\frac{(\rho_1+\rho_2)^2-1}{\sin^2(k\theta)}+\frac{(\rho_1-\rho_2)^2-1}{\cos^2(k\theta)}\right]
\end{equation}
Now, \eqref{eq:system1} for \(j=3\) has
\begin{equation}
4k^2c_3       =-4kc_1V\cos(2k\theta)-c_1\der{}{V}{\theta}\sin(2k\theta)\label{eq:j3}
\end{equation}
So \(c_1=k\) giving
\begin{equation}
J=\left[\sin(2k\theta)\partial_\theta+\cos(2k\theta)N\right]\left(N+k\right)+k^2\rho_1\rho_2\label{eq:2ord}
\end{equation}
The norm polynomial is
\begin{equation}
    \Phi\left(N\right)=\left[\left(N-k\right)^2-k^2\rho_1^2\right]\left[\left(N-k\right)^2-k^2\rho_2^2\right]
\end{equation}
which is of the form \eqref{eq:normpol}. We obtain the structure relation
\begin{equation}
    J_+J_-+J_-J_+-2M^2+2k^2M(\rho_1^2+\rho_2^2-6)=2k^4(\rho_1^2-1)(\rho_2^2-1)
\end{equation}
Set \(\alpha=(\rho_1+\rho_2)/2\) and \(\beta=(\rho_1-\rho_2)/2\). The eigenfunctions are given by
\[\Theta_n(\theta)=[\sin(k\theta)]^{\frac{1}{2}+\alpha}[\cos(k\theta)]^{\frac{1}{2}+\beta}\mathscr{P}_n^{(\alpha,\beta)}(\cos(2k\theta)).\]
Here \(\alpha,\beta\) are Jacobi indices local to this section, not the PVI parameters used in Section~6. The \(\mathscr{P}_n^{(\alpha,\beta)}\) are Jacobi polynomials:
\begin{equation}
    \mathscr{P}^{(\alpha,\beta)}_n(x)=\frac{(-1)^n}{2^nn!}(1-x)^{-\alpha}(1+x)^{-\beta}\frac{\mathrm{d}^n}{\mathrm{d}x^n}\left[(1-x)^\alpha(1+x)^\beta\left(1-x^2\right)^n\right]
\end{equation}
The \(\Theta_n\) are square-integrable on \(\left(0,\frac{\pi}{2k}\right)\) for \(\alpha,\beta>-1\), and for \(1+\alpha+\beta>0\) the branch of \(N\) fixed in Section~\ref{sec:formal-ladders} is the positive square root of \(M\). We have the action
\begin{align*}
    M\Theta_n&=k^2(2n+1+\alpha+\beta)^2\Theta_n\\
    J_+\Theta_n&=4k^2(n+1)(n+\alpha+\beta+1)\Theta_{n+1}\\
    J_-\Theta_n&=\begin{cases}4k^2(n+\alpha)(n+\beta)\Theta_{n-1},&\quad n>0;\\0,&\quad n=0.\end{cases}
\end{align*}
By direct calculation, we find
\begin{align*}
    \phi^{(1)}&=k^2\rho_1\rho_2,\\
    \phi^{(2)}&=k^4\left(\rho_1^2-1\right)\left(\rho_2^2-1\right)
\end{align*}
where for the second we use
\begin{equation}
\begin{aligned}
J^2&=\{[\sin(4k\theta)\partial_\theta+\cos(4k\theta)N](N+k)(N+3k)\\
&\qquad+2k^2\rho_1\rho_2[\sin(2k\theta)\partial_\theta+\cos(2k\theta)(2N+3k)]\\
&\qquad+k^2(\rho_1^2+\rho_2^2-1)(N+2k)\}(N+2k)\\&\qquad+k^4(\rho_1^2-1)(\rho_2^2-1)
\end{aligned}
\end{equation}
Hence \(\phi^{(n)}\) is of the form \eqref{eq:phiform} for all \(n\), here taking \(\mu=2\).
\subsection{Single reduction of order}

\label{subsec:explicit-reduction}
Set \(\rho_1=m\in\mathbb N_0\) and keep \(\rho_2=\rho\) free. For generic \(\rho\), one norm parameter is an integer and gives one family of removable spectral factors. The corresponding Hamiltonians and intertwiners are
\begin{align*}
M_m&=-\partial_\theta^2+\frac{k^2(\rho^2+m^2-1)+2k^2m\rho\cos(2k\theta)}{\sin^2(2k\theta)},\\
\mathcal{A}_m&=\partial_\theta-k\rho\csc(2k\theta)-k(m+1)\cot(2k\theta)
\end{align*}
The shift \(m\mapsto m+2\) separates the even and odd values of \(m\) into two Darboux chains. The factorization relations are
\begin{align*}
    \mathcal{A}_mM_m&=M_{m+2}\mathcal{A}_m\\
    \mathcal{A}_m^\dagger \mathcal{A}_m&
    =M_m-k^2(m+1)^2\\
    \mathcal{A}_m\mathcal{A}_m^\dagger&=M_{m+2}-k^2(m+1)^2
\end{align*}
Dressing preserves the step \(2nk\) and adds two to the formal order. For generic \(\rho\), with the right-factor normalization used above, the reduced operators satisfy
\[L^{(n)}_{m+2}=\mathcal{A}_mL^{(n)}_m\mathcal{A}_m^\dagger,\qquad n\geq m+2.\]
In particular, for \(m=0\), we have a first-order ladder operator
\[L^{(1)}_0=\sin(2k\theta)\partial_\theta+\cos(2k\theta)N_0.\]
Starting from this first-order seed gives the even chain:
\[
 L^{(n)}_{2m}=\left(\prod^{m}_{i=1}\mathcal{A}_{2m-2i}\right)
 \left[L^{(1)}_0\right]^n
 \left(\prod^{m-1}_{i=0}\mathcal{A}_{2i}^\dagger\right),
 \qquad n\geq2m.
\]
The right-hand product is the adjoint of the left-hand Darboux chain. Its indices must therefore be \(0,2,\ldots,2m-2\). The dressed expression has formal order \(2m+n\), compared with \(2n\) for \(J^n\), so a strict reduction occurs for \(n>2m\). The odd chain starts at \(M_1\) and requires a separate seed. It is not included in this even-chain formula.
%
\subsection{Double reduction of order}
Now take both norm parameters to be integers of opposite parity,
\(\rho_1=2m\) and \(\rho_2=2n-1\), with \(m\geq0\) and \(n\geq1\). There is one family of removable factors for each parity. The following intertwiners change the two chain indices independently:
\begin{align*}
    M_{m,n}&=-\partial_\theta^2+\frac{4k^2[m^2+n(n-1)]+4k^2m(2n-1)\cos(2k\theta)}{\sin^2(2k\theta)}\\
    \mathcal{A}_{m,n}&=\partial_\theta-k(2m+1)\cot(2k\theta)-k(2n-1)\csc(2k\theta)\\
    \mathcal{B}_{m,n}&=\partial_\theta-2kn\cot(2k\theta)-2km\csc(2k\theta)
\end{align*}
\begin{align*}
    \mathcal{A}_{m,n}M_{m,n}=M_{m+1,n}\mathcal{A}_{m,n}\\
    \mathcal{B}_{m,n}M_{m,n}=M_{m,n+1}\mathcal{B}_{m,n}
\end{align*}

Since \(M_{0,1}=-\partial_\theta^2\), we can first increase the second index from \(1\) to \(n\), and then the first index from \(0\) to \(m\). Define the forward chain
\[
 \mathcal C_{m,n}=
 \left(\prod^{m-1}_{i=0}\mathcal A_{m-1-i,n}\right)
 \left(\prod^{n-1}_{i=1}\mathcal B_{0,n-i}\right),
 \qquad
 \mathcal C_{m,n}M_{0,1}=M_{m,n}\mathcal C_{m,n},
\]
with empty products equal to the identity. The dressed ladder is therefore
\[
 L^{(\alpha)}_{m,n}=\mathcal C_{m,n}
 [\sin(2\alpha k\theta)\partial_\theta+
       \cos(2\alpha k\theta)N_{0,1}]
 \mathcal C_{m,n}^{\dagger}.
\]
The adjoint chain acts first, taking a state of \(M_{m,n}\) back to the free problem. Then the free ladder shifts it, and the forward chain returns it to \(M_{m,n}\). On the common transported eigenfunction span,
\[
 [N_{m,n},L^{(\alpha)}_{m,n}]=2\alpha kL^{(\alpha)}_{m,n},
 \qquad N_{m,n}=\sqrt{M_{m,n}}.
\]
Before any further factors are removed, its formal order is
\(2(m+n-1)+1=2m+2n-1\), independent of \(\alpha\). On the free Dirichlet basis, \(\alpha\) is a positive integer. This fixed-order construction contrasts with the single-reduction chain of Section~\ref{subsec:explicit-reduction}, where the order \(2m+h\) of the reduced \(h\)th power grows with \(h\).

\section{Third- and Fourth-Order Ladder Operators}
The purpose of this section is to extend this construction to higher order, by which we mean beyond second order. In many places in the context of integrability, superintegrability and various algebraic methods in quantum mechanics, there is a qualitative change when one goes beyond second order. In many contexts this comes from compatibility equations, which can then be nonlinear. This is again a key point in our approach, and it highlights how, in a general manner, transcendental functions such as the Painlev\'e transcendents are related to higher-order operators, in our case higher-order generalized ladder operators. In this section, we consider explicitly the case where the generating ladder operators are of third- or fourth-order, and cannot be reduced to the cases discussed in the foregoing sections.

We again assume \(F_0,F_1,G_0,G_1\) are defined as in \eqref{eq:a0}--\eqref{eq:a1}. For \(j=2\), \eqref{eq:system1}--\eqref{eq:system2} reads
\begin{align}
    4kF_2&=2\der{}{G_2}{\theta}-4kV\cos(2k\theta)-\der{}{V}{\theta}\sin(2k\theta)\\
    4kG_2&=-2\der{}{F_2}{\theta}\label{eq:b2eq}
\end{align}
By eliminating \(G_2\), we obtain a second-order equation for \(F_2\):
\begin{equation}
\der{2}{F_2}{\theta}+4k^2F_2=-4k^2V\cos(2k\theta)-k\der{}{V}{\theta}\sin(2k\theta)
\label{eq:vprime}\end{equation}There is a first integral
\begin{equation}\sin(2k\theta)\der{}{F_2}{\theta}-2k\cos(2k\theta)F_2=c_3-kV\sin^2(2k\theta)\label{eq:v}
\end{equation}
There are five more equations to solve from \eqref{eq:system1}--\eqref{eq:system2}:
\begin{align}
    4kF_3+4k^2F_2&=-\der{2}{F_2}{\theta}+2\der{}{G_3}{\theta}-2V\der{}{G_1}{\theta}-\der{}{V}{\theta}G_1\label{eq:3a}\\
    4kG_3+4k^2G_2&=-\der{2}{G_2}{\theta}-2\der{}{F_3}{\theta}\label{eq:3b}\\
            4kF_4+4k^2F_3&=-\der{2}{F_3}{\theta}-2V\der{}{G_2}{\theta}-\der{}{V}{\theta}G_2\label{eq:3c}\\
    4k^2G_3&=-\der{2}{G_3}{\theta}-2\der{}{F_4}{\theta}\label{eq:3d}\\
    4k^2F_4&=-\der{2}{F_4}{\theta}-2V\der{}{G_3}{\theta}-\der{}{V}{\theta}G_3\label{eq:3e}
\end{align}
From \eqref{eq:3b} and \eqref{eq:b2eq} we determine
\begin{equation}
    G_3=\frac{1}{2}\der{}{F_2}{\theta}+\frac{1}{8k^2}\der{3}{F_2}{\theta}-\frac{1}{2k}\der{}{F_3}{\theta}\label{eq:b3eq}
    \end{equation}
We substitute \eqref{eq:b3eq} into \eqref{eq:3a} to find
\begin{equation}
    \der{2}{F_3}{\theta}+4k^2F_3=\frac{1}{4k}\der{4}{F_2}{\theta}-4k^3F_2-2kV\der{}{G_1}{\theta}-kG_1\der{}{V}{\theta}\label{eq:a3eq}
\end{equation}
Integrating, we get
    \begin{equation}F_3=\frac{1}{4k}\der{2}{F_2}{\theta}+(c_1-k)F_2+c_4\cos(2k\theta)+c_5\sin(2k\theta)\label{eq:a3solve}
\end{equation}
We are thus able to solve \eqref{eq:3d}:
\begin{equation}
\begin{aligned}
F_4&=(c_1-2k)\left(\frac{1}{4k}\der{2}{F_2}{\theta}+kF_2\right)+c_6.
\end{aligned}\label{eq:a4solve}
\end{equation}
%
We substitute \eqref{eq:b2eq}, \eqref{eq:b3eq}, \eqref{eq:a3solve} and \eqref{eq:a4solve} into \eqref{eq:3c} and \eqref{eq:3e}. We obtain two fourth-order differential equations in \(F_2\) (after rescaling):
\begin{equation}
    0=\der{4}{F_2}{\theta}+(8kc_1-8k^2-4V)\der{2}{F_2}{\theta}-2\der{}{V}{\theta}\der{}{F_2}{\theta}+16k^3(2c_1-3k)F_2+16k^2c_6\label{eq:a41}
\end{equation}
\begin{equation}
\begin{aligned}
    0&=(c_1-2k)\left[\der{4}{F_2}{\theta}+4(2k^2-V)\der{2}{F_2}{\theta}-2\der{}{V}{\theta}\der{}{F_2}{\theta}+16k^4F_2\right]\\
    &\qquad+4k[c_4\sin(2k\theta)-c_5\cos(2k\theta)]\der{}{V}{\theta}\\
    &\qquad+16k^2[c_4\cos(2k\theta)+c_5\sin(2k\theta)]V+16k^3c_6\label{eq:a42}
\end{aligned}
\end{equation}
Eliminating \(\displaystyle\der{2}{F_2}{\theta}, \der{4}{F_2}{\theta}\) using \eqref{eq:vprime}, \eqref{eq:a41} and \eqref{eq:a42}, we get
\begin{equation}
    \begin{aligned}
    4kc_6(c_1-3k)&=\{[c_4+2k(c_1-2k)^2]\sin(2k\theta)-c_5\cos(2k\theta)\}\der{}{V}{\theta}\\
    &\qquad+4k\{[c_4+2k(c_1-2k)^2]\cos(2k\theta)+c_5\sin(2k\theta)\}V
\end{aligned}
\end{equation}
If this equation is non-trivial, then \(V\) reduces to a translated P\"{o}schl-Teller model. Therefore \(c_4=-2k(c_1-2k)^2,c_5=0\) and \(c_6\times(c_1-3k)=0\). Combining our formulas for \(F_i,G_i\) in terms of \(F_2\), we observe the identities:
\begin{align}
    \sum^4_{i=0}F_i(\theta)(2k-c_1)^{4-i}&=c_6\label{eq:fgid1}\\
    \sum^3_{i=0}G_i(\theta)(2k-c_1)^{3-i}&=0\label{eq:fgid2}
\end{align}
Now, if \(\mu=3\) then \(F_4=0\). If, in addition, \(c_1\neq 2k\) then from \eqref{eq:a4solve}, \(F_2\) is a trigonometric function and we have a P\"oschl-Teller potential again. So \(\mu=3\) implies \(c_1=2k,c_6=0\). 

Conversely, if \(\mu=4\) and \(c_6=0\), then from \eqref{eq:fgid1}--\eqref{eq:fgid2}, \(J\) admits a right-factor of \(N+c_1-2k\) and can be reduced to a third-order operator. Then if we set \(F_2\mapsto F_2+2k(c_1-2k)\cos(2k\theta)\), \(c_3\mapsto c_3-4k^2(c_1-2k)\) then \(V\) as defined in \eqref{eq:v} is unchanged, however \(c_1\mapsto 2k\) in \eqref{eq:a41}. In this way, we obtain the general potential corresponding to a third-order ladder operator as the special case of the fourth-order model by simply making the assignment \(c_6\mapsto 0\) (the value of \(c_1\) not being instrumental to the determination of \(V\)).

With this in mind, we fix \(\mu=4\) and \(c_1=3k\). After eliminating \(V,V'\), \eqref{eq:a41} becomes
\begin{equation}
\begin{aligned}
0&=\der{4}{F_2}{\theta}+\frac{6}{k\sin(2k\theta)}\der{}{F_2}{\theta}\der{2}{F_2}{\theta}-\frac{8F_2\cos(2k\theta)}{\sin^2(2k\theta)}\der{2}{F_2}{\theta}+\der{2}{F_2}{\theta}\left[16k^2-\frac{4c_3}{k\sin^2(2k\theta)}\right]\\
&\qquad-\frac{8\cos(2k\theta)}{\sin^2(2k\theta)}\left(\der{}{F_2}{\theta}\right)^2+8kF_2\der{}{F_2}{\theta}\left[\frac{2}{\sin^3(2k\theta)}-\frac{1}{\sin(2k\theta)}\right]+\frac{8c_3\cos(2k\theta)}{\sin^3(2k\theta)}\der{}{F_2}{\theta}\\
&\qquad+48k^4F_2+16k^2c_6
\end{aligned}
\label{eq:fourthorder}
\end{equation}
Integrating equation \eqref{eq:fourthorder} gives
\begin{equation}
\begin{aligned}
    0&=\sin(2k\theta)\der{3}{F_2}{\theta}-2k\cos(2k\theta)\der{2}{F_2}{\theta}+\frac{3}{k}\left(\der{}{F_2}{\theta}\right)^2-\frac{8F_2\cos(2k\theta)}{\sin(2k\theta)}\der{}{F_2}{\theta}-4kF_2^2\\
    &\qquad+4\der{}{F_2}{\theta}\left[3k^2\sin(2k\theta)-\frac{c_3}{k}\csc(2k\theta)\right]-24k^3F_2\cos(2k\theta)-8kc_6\cos(2k\theta)+c_7
\end{aligned}\label{eq:thirdorder}
\end{equation}
There is a second integral
\begin{equation}
\begin{aligned}
0&=\left(\der{2}{F_2}{\theta}\right)^2-\frac{4k\cos(2k\theta)}{\sin(2k\theta)}\der{}{F_2}{\theta}\der{2}{F_2}{\theta}+\frac{2}{k\sin(2k\theta)}\left(\der{}{F_2}{\theta}\right)^3-\frac{8F_2\cos(2k\theta)}{\sin^2(2k\theta)}\left(\der{}{F_2}{\theta}\right)^2\\
&\qquad-\frac{8kF_2^2}{\sin(2k\theta)}\der{}{F_2}{\theta}+\left(\der{}{F_2}{\theta}\right)^2\left[8k^2-\frac{4c_3}{k\sin^2(2k\theta)}\right]-\frac{48k^3F_2\cos(2k\theta)}{\sin(2k\theta)}\der{}{F_2}{\theta}-48k^4F_2^2\\
&\qquad+\frac{2c_7-16kc_6\cos(2k\theta)}{\sin(2k\theta)}\der{}{F_2}{\theta}-32k^2c_6F_2+c_8
    \end{aligned}\label{eq:secondorder}
\end{equation}
We make the substitution \(x=\cos^2(k\theta),F_2=8k^2u(x)+6k^2x-3k^2\). The ladder operator now takes the form
\begin{equation}
\begin{aligned}
J&=c_6+(N+k)\left\{(2x-1)\left[N(N+k)^2-4k^3\der{}{u}{x}\right]+8k^2(N+k)u\right.\\
&\qquad\left.-8k^3x(x-1)\der{2}{u}{x}+4kx(x-1)\left[\frac{\partial}{\partial x}(N+k)^2+4k^2\der{}{u}{x}\frac{\partial}{\partial x}\right]\right\}
\end{aligned}
\end{equation}
and equation \eqref{eq:secondorder} transforms into equation SD-Ia in Cosgrove and Scoufis' catalogue \cite{cosgrove91}:
\begin{equation}
\begin{aligned}
0&=x^2(x-1)^2\left(\der{2}{u}{x}\right)^2+4\der{}{u}{x}\left(x\der{}{u}{x}-u\right)^2-4\left(\der{}{u}{x}\right)^2\left(x\der{}{u}{x}-u\right)-\left(\frac{7}{4}+\frac{c_3}{4k^3}\right)\left(\der{}{u}{x}\right)^2\\
&\qquad+\frac{c_6}{4k^4}\left(x\der{}{u}{x}-u\right)-\left(\frac{15}{16}+\frac{3c_3}{8k^3}+\frac{c_6}{8k^4}+\frac{c_7}{64k^5}\right)\der{}{u}{x}-\frac{9}{64}-\frac{9c_3}{64k^3}-\frac{3c_7}{256k^5}+\frac{c_8}{1024k^8}
\end{aligned}
\end{equation}
It is solved by the substitution
\begin{align*}
    u&=\frac{x^2(x-1)^2}{4y(y-1)(y-x)}\left[\der{}{y}{x}-\frac{y(y-1)}{x(x-1)}\right]^2\\
    &\qquad+\tfrac{1}{32}(\rho_1+\rho_2)^2(1-2y)+\tfrac{1}{32}(\rho_1-\rho_2)^2\left[1-\frac{2x(y-1)}{y-x}\right]\\
    &\qquad+\tfrac{1}{32}(\rho_3+\rho_4)^2\left(1-\frac{2x}{y}\right)-\tfrac{1}{32}(\rho_3-\rho_4)^2\left[1-\frac{2(x-1)}{y-1}\right]\\
    c_3&=k^3\left(\rho_1^2+\rho_2^2+\rho_3^2+\rho_4^2-7\right)\\
   c_6&=k^4\rho_1\rho_2\rho_3\rho_4\\
    c_7&=4 k^5 \left[\rho_1^2\rho_2^2+\rho_1^2\rho_3^2+\rho_1^2\rho_4^2+\rho_2^2\rho_3^2+\rho_2^2\rho_4^2+\rho_3^2\rho_4^2\right.\\
    &\qquad\left.-6\left(\rho_1^2+\rho_2^2+\rho_3^2+\rho_4^2\right)+27\right]\\
 c_8&=16k^8\left[27-9\left(\rho_1^2+\rho_2^2+\rho_3^2+\rho_4^2\right)\right.\\
 &\qquad+3\left(\rho_1^2\rho_2^2+\rho_1^2\rho_3^2+\rho_1^2\rho_4^2+\rho_2^2\rho_3^2+\rho_2^2\rho_4^2+\rho_3^2\rho_4^2\right)\\
 &\qquad\left.-\left(\rho_1^2\rho_2^2\rho_3^2+\rho_2^2\rho_3^2\rho_4^2+\rho_1^2\rho_3^2\rho_4^2+\rho_1^2\rho_2^2\rho_4^2\right)\right]
\end{align*}
where \(y=\mathrm{P}_{\mathrm{VI}}(x;\alpha,\beta,\gamma,\delta)\) is the sixth Painlev\'{e} transcendent, with the conventional parameters related to \(\rho_j\) by \eqref{eq:concise-parameter-dictionary}. In these variables it satisfies
\begin{equation}
\begin{aligned}
\der{2}{y}{x}&=\frac{1}{2}\left(\frac{1}{y}+\frac{1}{y-1}+\frac{1}{y-x}\right)\left(\der{}{y}{x}\right)^2-\left(\frac{1}{x}+\frac{1}{x-1}+\frac{1}{y-x}\right)\der{}{y}{x}\\
&\qquad+\frac{y(y-1)(y-x)}{8x^2(x-1)^2}\left[(\rho_1+\rho_2+2)^2-\frac{(\rho_3+\rho_4)^2x}{y^2}+\frac{(\rho_3-\rho_4)^2(x-1)}{(y-1)^2}\right.\\
&\qquad\left.-\frac{(\rho_1-\rho_2+2)(\rho_1-\rho_2-2)x(x-1)}{(y-x)^2}\right].
\end{aligned}
\end{equation}
We arrive at the potential
\begin{equation}
\begin{aligned}
V&=\frac{k^2x(x-1)(2y-1)}{y(y-1)(x-y)}\left\{\der{}{y}{x}-\frac{y(y-1)[2y-1-(\rho_1+\rho_2)(x-y)]}{x(x-1)(2y-1)}\right\}^2+\frac{k^2(\rho_1+\rho_2)^2(x-y)}{4x(x-1)(2y-1)}\\
&\qquad+\frac{k^2(\rho_3+\rho_4)^2(y-1)}{4y(x-1)}+\frac{k^2(\rho_3-\rho_4)^2y}{4x(1-y)}+\frac{k^2(\rho_1-\rho_2)^2(2y-1)}{4(y-x)}+\frac{k^2}{4x(x-1)}
\end{aligned}
\end{equation}
We may determine \(\Phi\) in terms of \(F_2\) and its derivatives after eliminating all of the other equations. After substituting equations \eqref{eq:fourthorder}, \eqref{eq:thirdorder} and \eqref{eq:secondorder}, \(F_2\) disappears completely and we have
\begin{equation}
\begin{aligned}
\Phi\left(N\right)&=\left[\left(N-k\right)^2-k^2\rho_1^2\right]\left[\left(N-k\right)^2-k^2\rho_2^2\right]\\
&\qquad\times\left[\left(N-k\right)^2-k^2\rho_3^2\right]\left[\left(N-k\right)^2-k^2\rho_4^2\right]
\end{aligned}
\end{equation}
which is of the form \eqref{eq:normpol} for \(\mu=4\). When \(\mu=3\), \(c_6=0\) and one of the \(\rho_i\), say \(\rho_4\), must be set to 0. Then \(J_\pm\) has the right factor \(k\pm N\), and the third-order operators are \(\widetilde{J}_\pm=J_\pm(k\pm N)^{-1}\). Moving \((k+N)^{-1}\) through \(J_-\) gives
\[\widetilde{\Phi}\left(N\right)=\widetilde J_+\widetilde J_-=-\frac{\Phi(N)}{(N-k)^2}=\prod^3_{j=1}\left[k^2\rho_j^2-\left(N-k\right)^2\right],\]
Thus \eqref{eq:normpol} holds for \(\mu=3\) as well.

The \(\phi\) parameters also satisfy \eqref{eq:phiform} in the fourth-order case. We calculate explicitly
\begin{align*}
    \phi^{(1)}&=k^4\rho_1\rho_2\rho_3\rho_4,\\
    \phi^{(2)}&=k^8(\rho_1^2-1)(\rho_2^2-1)(\rho_3^2-1)(\rho_4^2-1)
\end{align*}
and obtain the rest by induction.

This completes the construction for ladder operators of order at most four. First- and second-order ladders give the P\"oschl--Teller family, whereas third- and fourth-order ladders lead to the Painlev\'e VI potential, with the norm polynomial~\eqref{eq:normpol} determined by the four parameters \(\rho_j\). So far, however, \(M\) has been treated as a one-dimensional operator. In the next section we view it as the angular part of a separable two-dimensional Hamiltonian and combine \(J_\pm\) with radial shift operators to obtain integrals of motion of the full system.

\section{Superintegrable Extensions of Quadratic Hamiltonians}
\label{sec:superintegrable}
We now combine the angular ladder operators with radial shift operators to construct integrals of motion of two-dimensional Hamiltonians that separate in polar coordinates. In such a Hamiltonian the angular operator enters the radial equation only through \(M=N^2\). The ladder operator \(J_+\) raises \(N\) by \(2k\), while a radial shift operator raises the label \(N\) of the radial Hamiltonian by \(2\ell\) at fixed energy. When \(k=p\ell/q\) with coprime integers \(p,q\), the product \(J_+^qK^{(p)}_+\) raises both by the same amount and therefore commutes with the Hamiltonian. We construct these integrals for the oscillator, Coulomb and free radial problems, compute their norm polynomials, and show that together with \(M\) they generate a deformed Racah-type polynomial algebra.

Consider the polar-separable Hamiltonian

\begin{equation}
 H_N=\frac12\left(-\partial_r^2-\frac1r\partial_r+\frac{N^2}{r^2}\right)+U(r).
 \label{eq:polar-HM}
\end{equation}
A one-step radial shift operator with parameter step $2\ell$ satisfies
\begin{equation}
 [K,H_N]=\frac{2\ell(N+\ell)}{r^2}K,
 \qquad\Longleftrightarrow\qquad H_{N+2\ell}K=KH_N.
 \label{eq:radial-shift}
\end{equation}
We introduce higher order shift operators as follows:
\begin{equation}
    K^{(1)}_\pm\coloneqq K|_{N\mapsto\pm N},\qquad K^{(n)}_\pm=K^{(n-1)}_\pm|_{N\mapsto N\pm2\ell}K^{(1)}_\pm
\end{equation}
If $k=p\ell/q$ with coprime positive integers $p,q$, then
\begin{equation}
 X_+=J_+^qK^{(p)}_+,\qquad X_-=J_-^qK^{(p)}_-
 \label{eq:Xpm-superintegrable}
\end{equation}
commute with the full Hamiltonian. The three radial models used here are summarized in Table~\ref{tab:radial-shifts}. In the Coulomb row, $\varrho=\sqrt{-2E}$, while $J_N$ in the free Bessel row denotes the Bessel function of the first kind.  The table records only the one-step shift relation~\eqref{eq:radial-shift}, and it is not intended as a classification of all possible radial intertwiners.
\begin{center}
\begin{table}[ht]
\centering
\caption{Radial Hamiltonians and one-step shift operators used in the construction.}
\label{tab:radial-shifts}
\footnotesize
\renewcommand{\arraystretch}{1.18}
\begin{tabularx}{\textwidth}{@{}l l >{\raggedright\arraybackslash}X >{\raggedright\arraybackslash}X c >{\raggedright\arraybackslash}X@{}}
\toprule
Model & $U(r)$ & Radial eigenstate & Spectrum & $\ell$ & Shift $K$\\
\midrule
SHO
& $\frac12\omega^2r^2$
& $r^N e^{-\omega r^2/2}\,\mathscr L_n^N(\omega r^2)$
& $E=\omega(2n+1+N)$
& $1$
& $H+(N+1)(r^{-1}\partial_r-Nr^{-2})$\\[1mm]
Coulomb
& $-\kappa/r$
& $r^N e^{-\varrho r}\,\mathscr L_n^{2N}(2\varrho r)$
& $E=-\kappa^2/[2(n+N+\frac12)^2]$
& $\frac12$
& $\kappa+(N+\frac12)(\partial_r-N/r)$\\[1mm]
Free Bessel
& $0$
& $J_N(\sqrt{2E}\,r)$
& Continuous, $E>0$
& $\frac12$
& $\partial_r-N/r$\\
\bottomrule
\end{tabularx}
\end{table}
\end{center}
Define the radial norm polynomial by
\begin{equation}
 \Psi^{(n)}(H,N)\coloneqq K^{(n)}_+|_{N \mapsto N-2n\ell}K^{(n)}_-
 \end{equation}
To determine this function explicitly, we need only determine \(\Psi^{(1)}\) and use the obvious recurrence relation
\begin{equation}
\Psi^{(n+1)}(H,N)=\Psi^{(n)}(H,N-2\ell)\Psi^{(1)}(H,N).
\end{equation}
Then
\begin{align}
 \Psi^{(n)}_{\rm SHO}&=\prod_{i=1}^{n}\left[H^2-\omega^2(N-2i+1)^2\right],\\
 \Psi^{(n)}_{\rm Coulomb}&=\prod_{i=1}^{n}\left[2\left(N-i+\tfrac12\right)^2H+\kappa^2\right],\\
 \Psi^{(n)}_{\rm free}&=(-2H)^n.
\end{align}
Let us introduce the two polynomials in $H$ and $M$
\begin{align}
 \mathcal F_{p,q}(H,M)&=\sqrt{M}^{-1}\left[\Phi^{(q)}\left(\sqrt{M}\right)\Psi^{(p)}\left(H,\sqrt{M}\right)-\Phi^{(q)}\left(-\sqrt{M}\right)\Psi^{(p)}\left(H,-\sqrt{M}\right)\right] \label{eq:FPQ-early}\\
 \mathcal G_{p,q}(H,M)&=\Phi^{(q)}\left(\sqrt{M}\right)\Psi^{(p)}\left(H,\sqrt{M}\right)+\Phi^{(q)}\left(-\sqrt{M}\right)\Psi^{(p)}\left(H,-\sqrt{M}\right)\label{eq:GPQ-early}
\end{align}
and the generators
\begin{equation}
\begin{aligned}
 A&=\frac{M}{4p^2\ell^2}-\frac14,\\
 B&=\left(X_+-X_-\right)N^{-1},\\
 C&=\frac1{p\ell}\left[X_+\left(1+p\ell N^{-1}\right)
 +X_-\left(1-p\ell N^{-1}\right)\right].
\end{aligned}
\label{eq:ABC-generators}
\end{equation}
After normal ordering, the apparent factors $N^{-1}$ cancel and both operators are polynomial in $M$. Hence they are actual differential integrals. Their algebra is
\begin{subequations}\label{eq:deformed-Racah-algebra}
\begin{align}
 [A,B]&=C,\\
 [A,C]&=2\{A,B\},\\
 [B,C]&=-2B^2
 +\frac{2}{p\ell}\mathcal F_{p,q}\left(H,4p^2\ell^2A+p^2\ell^2\right),
 \label{eq:BC-deformed-Racah}\\
 0&=C^2+4B^2-2\{A,B^2\}\\
 &\qquad-2p^{-1}\ell^{-1}\mathcal F_{p,q}\left(H,4p^2\ell^2A+p^2\ell^2\right)\\
 &\qquad-2p^{-2}\ell^{-2}\mathcal G_{p,q}\left(H,4p^2\ell^2A+p^2\ell^2\right)
\end{align}
\end{subequations}
where \(\{u,v\}=uv+vu\).
%
%
\subsection{An explicit quadratic-algebra realization}
\label{subsec:explicit-quadratic-algebra}
To display the polynomial algebra in a completely explicit form, consider the oscillator radial problem and the first-order angular ladder with \(p=q=\ell=1\).  That is,
\begin{equation}
H=\frac{1}{2}\left[-\frac{\partial^2}{\partial r^2}-\frac{1}{r}\frac{\partial}{\partial r}-\frac{1}{r^2}\frac{\partial^2}{\partial\theta^2}+\frac{\rho^2-1}{r^2\sin^2(2\theta)}+ \omega^2r^2\right]
\end{equation}
Here \(\rho\) is the same angular parameter as in Section~\ref{firstorder}. Up to a common nonzero normalization of the composite integrals $X_\pm$, the angular and radial norm factors are
\begin{equation}
 \Phi(N)=\rho^2-(N-1)^2,
 \qquad
 \Psi(H,N)=H^2-\omega^2(N-1)^2.
\end{equation}
Hence
\begin{align}
 \mathcal F_{1,1}(H,M)&=4H^2+8\omega^2(\tfrac12\rho^2-M-1)\\
 \mathcal G_{1,1}(H,M)&=-2H^2(M+1-\rho^2)+2\omega^2(M^2+6M+1)-2\rho^2\omega^2(M+1)
\end{align}
For the present normalization, $M=4A+1$.  Substitution of
the above expressions for $\mathcal F_{1,1}$ and $\mathcal G_{1,1}$ into Eq.~\eqref{eq:deformed-Racah-algebra}
therefore gives
\begin{equation}
 \begin{aligned}
 [A,B]&=C,\\
 [A,C]&=2\{A,B\},\\
 [B,C]&=-2B^2+8H^2+32\omega^2(\tfrac14\rho^2-2A-1)\\
 0&=C^2+4B^2-2\{A,B^2\}+16H^2(A-\tfrac14\rho^2)-64\omega^2A(A+1-\tfrac14\rho^2)
 \end{aligned}
 \label{eq:explicit-quadratic-algebra}
\end{equation}
The Hamiltonian $H$ is central, so on each energy eigenspace \eqref{eq:explicit-quadratic-algebra} is a quadratic algebra with numerical structure constants. For $\rho>-1$ the oscillator levels are $E_{\mathcal N}=\omega(2\mathcal N+\rho+2)$, $\mathcal N=0,1,2,\ldots$, and on each level the algebra acts irreducibly on the $\mathcal N+1$ degenerate states with angular labels $1+\rho+2j$, $0\le j\le\mathcal N$.

Equivalently, the last relation of
\eqref{eq:explicit-quadratic-algebra} gives the Casimir value
\begin{align*}
 \mathscr K={}&C^2-2\{A,B^2\}+4B^2-64\omega^2A^2\\
 &+16\bigl[H^2+\omega^2(\rho^2-4)\bigr]A
 =4\rho^2H^2.
\end{align*}
For \(\omega>0\) and \(\rho>-1\), choose the deformed-oscillator
sign convention with structure function
\(-X_+X_-\). On the level \(E_{\mathcal N}\), evaluation at
\(N=1+\rho+2j\) gives
\[
 \varphi_{\mathcal N}(j)
 =16\omega^2j(j+\rho)(\mathcal N-j+1)
                 (\mathcal N+j+\rho+1).
\]
It vanishes at \(j=0\) and \(j=\mathcal N+1\) and is positive for
\(1\le j\le\mathcal N\), recovering the
\((\mathcal N+1)\)-dimensional multiplets.

The same mechanism governs the Painlev\'e VI examples of Sections~\ref{sec:PVI-special-solutions} and~\ref{sec:exceptional-Jacobi-Gromak}. The integrals $X_\pm$ built in this section are, however, not always of minimal differential order, as we show next.

\section{Reduction of Order}
\label{sec:reduction-order}
The question of constructing symmetries of minimal order is important from different perspectives. First, it is an important characteristic of superintegrability: not only the number of integrals matters, but also the order of all the integrals present for a given Hamiltonian. Finding integrals of low order helps describe the symmetry algebra. However, determining whether all independent symmetry generators have been found requires a further analysis of the Hamiltonian. This comes with a certain number of remarks. In certain cases associated with exceptional orthogonal polynomials, the integrals of minimal order may not describe the degeneracies completely, and another set of integrals, even though of higher order, allows one to obtain the spectrum completely and in an algebraic manner. This points out that, although finding symmetries of minimal order is important, there are additional aspects. It is, however, in general a difficult problem to establish whether integrals of minimal order have been obtained for a given Hamiltonian. An advantage of the method developed in this paper is that it allows a systematic treatment and the application of reduction of order. The commuting products of the preceding section need not have minimal differential order.  For the radial part define
\begin{equation}
 \psi^{(n)}(H)= K^{(n)}_+|_{N=-n\ell}. 
\end{equation}
For the oscillator and Coulomb shifts this is a polynomial in $H$, so the factor theorem gives
\begin{equation}
 \frac{K^{(n)}_+ -\psi^{(n)}(H)}{N+n\ell}
 \quad\text{polynomial in } N.
 \label{eq:radial-remainder-factor}
\end{equation}
The recurrence
\begin{equation}
 \psi^{(n+2)}(H)=\psi^{(n)}(H)\Psi^{(1)}(H,(n+2)\ell)
\end{equation}
with $\psi^{(0)}=1$, $\psi^{(1)}_{\rm SHO}=H$, and $\psi^{(1)}_{\rm K}=\kappa$ yields
\begin{align}
 \psi^{(2m)}_{\rm SHO}&=\prod_{i=1}^{m}[H^2-\omega^2(2i-1)^2],\\
 \psi^{(2m+1)}_{\rm SHO}&=H\prod_{i=1}^{m}[H^2-4i^2\omega^2],\label{eq:psi-SHO-corrected}\\
 \psi^{(2m)}_{\rm K}&=\prod_{i=1}^{m}[2(i-\tfrac12)^2H+\kappa^2],\\
 \psi^{(2m+1)}_{\rm K}&=\kappa\prod_{i=1}^{m}[2i^2H+\kappa^2].\label{eq:psi-K-corrected}
\end{align}
These formulas satisfy
\begin{equation}
 \Psi^{(n)}(H,n\ell)=[\psi^{(n)}(H)]^2.
 \label{eq:Psi-psi-square}
\end{equation}

The angular analogue is $\phi^{(n)}=J_+^n|_{N=-nk}$.  Since $qk=p\ell$, the formal integral $X_+=J_+^qK^{(p)}_+$ has the special value
\begin{equation}
 \chi_{p,q}(H)=X_+|_{N=-p\ell}=\phi^{(q)}\psi^{(p)}(H).
\end{equation}
Consequently,
\begin{equation}
 Y_\pm=[X_\pm -\chi_{p,q}(H)](p\ell\pm N)^{-1}
 \label{eq:reduced-Xpm}
\end{equation}
are again commuting operators and are polynomial in $N$ whenever $\phi^{(q)}$ is constant for the chosen power $q$ and the radial remainder property \eqref{eq:radial-remainder-factor} holds. Define the reduced generators below, with every inverse spectral factor acting on the right. The expressions involving \(A^{-1}\) are rewritings of the polynomial constructions in \(Y_\pm\), not independent choices of an inverse on a zero eigenspace:
\begin{align*}
    U&=A\\
    V&=p\ell(Y_+-Y_-)N^{-1}=[C-2B-2p^{-1}\ell^{-1}\chi_{p,q}(H)](4 A)^{-1}\\
    W&=B
\end{align*}
They satisfy
\begin{align*}
    [U,V]&=W\\
    [U,W]&=2\{U,V\}+2p^{-1}\ell^{-1}\chi_{p,q}(H)\\
    [V,W]&=-2V^2-\tfrac{1}{4}p^{-1}\ell^{-1}(2U^{-1}+U^{-2})\mathcal{F}_{p,q}(H,4p^2\ell^2U+p^2\ell^2)\\
    &\qquad-\tfrac{1}{4}p^{-2}\ell^{-2}U^{-2}\mathcal{G}_{p,q}(H,4p^2\ell^2U+p^2\ell^2)+\tfrac{1}{2}p^{-2}\ell^{-2}U^{-2}[\chi_{p,q}(H)]^2\\
    0&=W^2+4V^2-4p^{-1}\ell^{-1}\chi_{p,q}(H)V-2\{U,V^2\}\\
    &\qquad+\tfrac{1}{2}p^{-1}\ell^{-1}(3U^{-1}+U^{-2})\mathcal{F}_{p,q}(H,4p^2\ell^2U+p^2\ell^2)\\
    &\qquad+\tfrac{1}{2}p^{-2}\ell^{-2}(U^{-1}+U^{-2})\mathcal{G}_{p,q}(H,4p^2\ell^2U+p^2\ell^2)\\
    &\qquad-p^{-2}\ell^{-2}(U^{-1}+U^{-2})[\chi_{p,q}(H)]^2
\end{align*}
The combinations
\begin{equation}
 \mathcal I_1=Y_++Y_-,\qquad
 \mathcal I_2=\frac{Y_+-Y_-}{N}
 \label{eq:reduced-actual-integrals}
\end{equation}
are actual differential integrals of lower order.  Further reductions can occur on the angular resonant loci described by Theorem~\ref{th:reduce}.  The free Bessel shift remains useful for the formal construction, but its special evaluation is not generally a polynomial solely in $H$, so the radial factorization \eqref{eq:radial-remainder-factor} does not apply unchanged.

Commensurability gives a commuting formal product, and the radial and angular remainder factors can then be used to reduce its differential order. The question of whether these reductions yield integrals of the lowest possible order requires a more detailed analysis of the particular Hamiltonian.

The construction so far holds for a generic Painlev\'e VI transcendent, but it is formal. The roots of the norm polynomial are only candidate lowest or highest weights, and whether they correspond to physical states depends on the domain, the boundary conditions and the behaviour of the transcendent on the interval. In the next section we make this concrete for the special solutions of Painlev\'e VI, for which the potential is explicit and the spectral problem can be solved.

\subsection{Explicit order reduction along a Darboux chain}
\label{subsec:darboux-chain-reduction}
Consider the integer-parameter subfamily of the first-order potential given by 
\eqref{eq:V1}, with \(k>0\):
\begin{equation}
 M_r=-\partial_\theta^2+
       \frac{4k^2r(r-1)}{\sin^2(2k\theta)},
 \qquad r=1,2,\ldots,\quad
 0<\theta<\frac{\pi}{2k}.
 \label{eq:darboux-PT-chain}
\end{equation}
Here \(M_1=-\partial_\theta^2\) is the free Dirichlet operator. A ground state of \(M_r\) is
\(\psi_r(\theta)= \sin^r(2k\theta)\), with energy
\(\varepsilon_r=(2kr)^2\).  Its first-order intertwiner is
\begin{equation}
 A_r=\partial_\theta-2kr\cot(2k\theta),\qquad
 A_r^\dagger A_r=M_r-(2kr)^2,\qquad
 A_rA_r^\dagger=M_{r+1}-(2kr)^2.
 \label{eq:darboux-PT-intertwiner}
\end{equation}
The potentials \(M_r\) are therefore obtained by successive
state-deleting Darboux transformations of the free particle.  The
Wronskian representation for this trigonometric P\"oschl-Teller
chain is a special case of the results of Gaillard and Matveev
\cite{gaillardmatveev09}. An alternative proof based on shape
invariance was also given by Grandati \cite{grandati14}.  With
\(\mathcal B_{r-1}=A_{r-1}\cdots A_1\), the corresponding Crum
formula takes the explicit form
\begin{equation}
 \begin{aligned}
 V_r&=-2\frac{d^2}{d\theta^2}
       \log W(\sin(2k\theta),\ldots,
                    \sin(2(r-1)k\theta)),\\
 W(\sin(2k\theta),\ldots,
                    \sin(2(r-1)k\theta))
   &\propto\sin^{r(r-1)/2}(2k\theta),\\
 \mathcal B_{r-1}^\dagger\mathcal B_{r-1}
   &=\prod_{j=1}^{r-1}[M_1-(2kj)^2].
 \end{aligned}
 \label{eq:darboux-PT-crum}
\end{equation}
The empty Wronskian and products are understood to equal one for \(r=1\). For the free Dirichlet eigenfunctions
\(f_m=\sin(2km\theta)\), \(m\geq1\), with
\(N_1f_m=2kmf_m\), introduce the raising operators
\begin{equation}
 T_h=\sin(2hk\theta)\partial_\theta
       +\cos(2hk\theta)N_1,\qquad h=1,2,\ldots.
 \label{eq:darboux-free-step-ladders}
\end{equation}
They act as \(T_hf_m=2km f_{m+h}\), and their powers satisfy the following relation
\begin{equation}
 T_1^{\,h}=T_h\prod_{j=1}^{h-1}(N_1+2kj),
 \label{eq:darboux-free-power}
\end{equation}
with the spectral factors on the right.  On the Dirichlet spectrum of \(M_r\), write \(N_r=\sqrt{M_r}\) and set
\begin{equation}
 S_r=\sin(2k\theta)\partial_\theta
       +\cos(2k\theta)N_r,
 \qquad
 \mathcal D_{r,h}=\mathcal B_{r-1}T_h\mathcal B_{r-1}^\dagger.
 \label{eq:darboux-PT-dressed-shifts}
\end{equation}
The dressed ladder \(\mathcal D_{r,h}\) has formal order \(2r-1\),
independent of the step \(2hk\).  Here ``formal order'' counts \(N_r\) as first order, a ladder containing \(\sqrt{M_r}\) need not itself be an ordinary differential operator.

Put \(g_m=\mathcal B_{r-1}f_m\) for \(m\geq r\).  The
Gegenbauer recurrence for these eigenfunctions and the Crum
factorization \eqref{eq:darboux-PT-crum} give, in this normalization,
\begin{equation}
 \begin{aligned}
 S_rg_m&=2k(m-r+1)\,g_{m+1},\\
 \mathcal D_{r,h}g_m
   &=(2k)^{2r-1}(m-r+1)_{2r-1}\,g_{m+h},
 \end{aligned}
 \label{eq:darboux-PT-intrinsic-action}
\end{equation}
where \((a)_j\) denotes the rising factorial.  Consequently,
\(S_r^{\,h}g_m=(2k)^h(m-r+1)_h g_{m+h}\).  The factorization
\((a)_h=(a)_{2r-1}(a+2r-1)_{h-2r+1}\) then yields
\begin{equation}
 S_r^{\,h}=\mathcal D_{r,h}
    \prod_{j=0}^{h-2r}\bigl[N_r+2k(r+j)\bigr],
 \qquad h\geq2r-1.
 \label{eq:darboux-PT-reduction-all}
\end{equation}
The product is one when \(h=2r-1\).  Thus arbitrarily high
powers of the intrinsic ladder factor through dressed free-particle
ladders of fixed formal order \(2r-1\).

For \(r=2\), with \(V_2=8k^2\csc^2(2k\theta)\), this gives
\begin{equation}
 \mathcal D_{2,3}=S_2^{\,3},\qquad
 S_2^{\,4}=\mathcal D_{2,4}(N_2+4k).
 \label{eq:darboux-PT-r2-examples}
\end{equation}
These are identities on the Dirichlet eigenfunction span, with
spectral factors acting on the right.  The formal lowering partners
must separately respect the nonnegative spectrum and the chosen
self-adjoint domain near the lowest states. The explanation of the reduction of order based on Darboux transformations concerns this special
P\"oschl-Teller chain. This does not apply to the generic Painlev\'e VI potential. For the generic Painlev\'e VI, zeroes of the norm polynomial are only candidate spectral thresholds until the self-adjoint realization and the physical ladder modules are determined.  In the next section we examine this distinction for explicit special solutions.

\section{Special Painlev\'e VI solutions and spectral classes}
\label{sec:PVI-special-solutions}

Painlev\'e VI is naturally formulated as an isomonodromic deformation problem, and its special solutions are organized by the affine $D_4^{(1)}$ symmetry and B\"acklund transformations \cite{jim82,okamoto87pvi,iwasaki91,grom02}.  Rational and algebraic solutions have been studied from complementary arithmetic and monodromy viewpoints \cite{mazz01,yuan02,boalch05,boalch06,lisovyy14}.  Here we use the explicit families in GLS Tables 48.1--48.5 only to extract the information relevant to the quantum problem: the induced angular potential, the zeros of the ladder norm, the self-adjoint domain and the resulting spectral module.

\subsection{Parameter branches and ladder thresholds}

We retain the conventional parameters \((\alpha,\beta,\gamma,\delta)\) of \(P_{\rm VI}\) and the norm parameters \(\rho_1,\ldots,\rho_4\) used in Section~6. Their relation is
\begin{equation}
\begin{aligned}
 \alpha&=\frac{(\rho_1+\rho_2+2)^2}{8},&
 \beta&=-\frac{(\rho_3+\rho_4)^2}{8},\\
 \gamma&=\frac{(\rho_3-\rho_4)^2}{8},&
 \delta&=\frac{4-(\rho_1-\rho_2)^2}{8}.
\end{aligned}
\label{eq:concise-parameter-dictionary}
\end{equation}
Thus
\begin{equation}
\begin{aligned}
 \rho_1+\rho_2&=-2\pm\sqrt{8\alpha},&
 (\rho_1-\rho_2)^2&=4-8\delta,\\
 (\rho_3+\rho_4)^2&=-8\beta,&
 (\rho_3-\rho_4)^2&=8\gamma.
\end{aligned}
\label{eq:concise-alpha-branches}
\end{equation}
The potential depends on \(\rho_1+\rho_2\) unsquared inside its first squared term, so its two branches can define inequivalent Hamiltonians. The other three combinations enter only through their squares. Their sign changes leave the potential unchanged and only permute the \(\rho_j\) up to signs. Equivalently, choosing the same sign in the first two expressions gives
\begin{equation}
\begin{aligned}
 \rho_1&=-1\pm\sqrt{2\alpha}+\sqrt{1-2\delta},&
 \rho_2&=-1\pm\sqrt{2\alpha}-\sqrt{1-2\delta},\\
 \rho_3&=\sqrt{-2\beta}+\sqrt{2\gamma},&
 \rho_4&=\sqrt{-2\beta}-\sqrt{2\gamma},
\end{aligned}
\end{equation}
where the signs of the remaining square roots may be chosen independently. Define

\begin{equation}
 F_\rho(N)=(N-k)^2-k^2\rho^2.
\end{equation}
The one-step angular norm is then
\begin{equation}
\Phi^{(1)}(N)=\prod_{j=1}^4F_{\rho_j}(N).
\label{eq:concise-Phi-PVI}
\end{equation}
The corresponding formal roots are only candidate lowest or highest weights.  A physical module additionally requires a regular differential expression on the chosen interval, the appropriate endpoint conditions, a normalizable annihilation state and positivity of the norm. Below, $\nu$ denotes an eigenvalue of $N$. For a commensurate two-dimensional extension, $k=p\ell/q$, the same angular factor enters the full deformed-oscillator product
\begin{equation}
 X_+X_-=\Phi^{(q)}(N)\Psi^{(p)}(H,N).
\label{eq:concise-full-structure-function}
\end{equation}

\subsection{Rational GLS families}

For the representative values used throughout the explicit catalogue,
\begin{equation}
 (h,\eta_1)=\left(3,\frac12\right),\qquad a=2,\qquad n_{\rm GLS}=3,
\end{equation}
all direct entries of Tables 48.1--48.3 reduce to the spectral classes summarized in Table~\ref{tab:concise-GLS-spectral-classes}. Here $m=0,1,2,\ldots$ in the spectral column. The quoted spectra use \(0<\theta<\pi/(2k)\) and the endpoint branches selected by the corresponding Jacobi or exceptional-Jacobi eigenfunctions. Where two local solutions are square-integrable, this branch choice is part of the boundary condition. At an endpoint with inverse-square coefficient \(-1/4\), the logarithmic solution is excluded.

\begin{table}[ht]
\centering
{\small
\renewcommand{\arraystretch}{1.10}
\setlength{\tabcolsep}{4.5pt}
\begin{tabular}{@{}c l l l@{}}
\toprule
GLS source & Potential classes & Positive $\nu/k$ & Representation type \\
\midrule
48.1 & $\mathcal P_3$ & $\frac32+2m$ & ordinary Jacobi \\
48.1 & $\mathcal P_1,\mathcal P_2$ & $\frac72+2m$ & ordinary Jacobi \\
48.1 & $\mathcal R_2,\mathcal R_3$ & $\frac72+2m$ & exceptional, isospectral \\
48.1 & $\mathcal P_4$ & $\frac{11}{2}+2m$ & ordinary Jacobi \\
48.1 & $\mathcal R_1$ & $\frac{11}{2}+2m$ & exceptional, isospectral \\
\midrule
48.2 & $\mathcal A_1$ & $3+2m$ & ordinary Jacobi \\
48.2 & $\mathcal A_2$ & $\{1\}\cup(5+2m)$ & added singlet $+$ ladder \\
48.2 & $\mathcal A_3$ & split domain & interior barrier \\
48.2 & $\mathcal A_4$ & $5+2m$ & ordinary Jacobi \\
48.2 & $\mathcal A_5$ & $\{3\}\cup(7+2m)$ & added singlet $+$ ladder \\
\midrule
48.3 & $\mathcal B_1,\mathcal B_3$ & $4+2m$ & ordinary Jacobi \\
48.3 & $\mathcal B_2,\mathcal B_4$ & $6+2m$ & exceptional, isospectral \\
48.3 & $\mathcal B_5$ & $6+2m$ & ordinary Jacobi\\
\bottomrule
\end{tabular}}
\caption{Spectral classes obtained from the representative rational GLS families.  Missing exceptional polynomial degrees need not correspond to missing physical energy levels.}
\label{tab:concise-GLS-spectral-classes}
\end{table}

Three examples display the essential phenomena.  Put $u=k\theta$ and $c=\cos^2u$.  Table 48.1 row 2 has
\begin{equation}
 y(x)=\frac{x(x+3)}{3(5-x)},\qquad
 (\alpha,\beta,\gamma,\delta)=\left(\frac98,-\frac18,2,-\frac32\right).
\end{equation}
The branch $\rho_1+\rho_2=1$ gives the ordinary P\"oschl--Teller potential
\begin{equation}
 \frac{V_+(\theta)}{k^2}=6\sec^2u-\frac14\csc^2u,
\end{equation}
whereas $\rho_1+\rho_2=-5$ gives the regular rational extension
\begin{equation}
 \frac{V_-(\theta)}{k^2}=6\sec^2u+\frac{15}{4}\csc^2u
 +\frac{36}{5-c}-\frac{160}{(5-c)^2}.
\label{eq:concise-R1-example}
\end{equation}
Thus a single conventional $P_{\rm VI}$ parameter point can lead to two distinct quantum Hamiltonians and two different lowest-weight lattices.

\paragraph{Polynomial algebra of the regular exceptional branch.}
For the regular branch in Eq.~\eqref{eq:concise-R1-example} choose
\begin{equation}
 (\rho_1,\rho_2,\rho_3,\rho_4)
 =\left(-\frac12,-\frac92,\frac52,-\frac32\right).
\end{equation}

For $k=1$, the four angular threshold parameters are then
$\{\tfrac12,\tfrac32,\tfrac52,\tfrac92\}$, and
\begin{equation}
 \Phi_{\mathcal R_1}(N)=
 \prod_{\rho\in\{1/2,3/2,5/2,9/2\}}
 \bigl[(N-1)^2-\rho^2\bigr].
 \label{eq:R1-angular-Phi}
\end{equation}
For the oscillator extension with $k=\ell=p=q=1$, the generators $A,B,C$
of Section~7 satisfy
\begin{equation}
 \begin{aligned}
 [A,B]&=C,\\
 [A,C]&=2\{A,B\},\\
 [B,C]&=-2B^2+\frac{1}{32}(16A-45)(16A-5)(16A+3)\\
 &\hspace{25mm}\times\bigl(80\omega^2A-16H^2+91\omega^2\bigr).
 \end{aligned}
 \label{eq:R1-explicit-polynomial-algebra}
\end{equation}

The principal angular ladder begins at $\nu_0=11/2$.  At fixed energy the
finite module is described by
\begin{equation}
 \nu_j=\frac{11}{2}+2j,
 \qquad
 E=\omega(2n+1+\nu_j)=\omega\left(2{\mathcal N}+\frac{13}{2}\right),
 \qquad {\mathcal N}=n+j,\quad 0\leq j\leq {\mathcal N},
\end{equation}
with positive structure function
\begin{equation}
 \begin{aligned}
 \varphi_{\mathcal N}^{(\mathcal R_1)}(j)
 ={}&32\omega^2j(j+1)(j+2)(j+3)
 (2j+3)(2j+5)(2j+7)(2j+9)\\
 &\times({\mathcal N}-j+1)(2{\mathcal N}+2j+11).
 \end{aligned}
 \label{eq:R1-deformed-oscillator}
\end{equation}
Thus the exceptional gaps in polynomial degree do not create an additional
spectral component: the two-dimensional polynomial algebra still acts
irreducibly on a single $({\mathcal N}+1)$-dimensional multiplet.

Table 48.2 contains both state addition and a change of self-adjoint domain.  With
\begin{equation}
 \Delta_2(u)=1-\frac12\sin^2(2u),\qquad
 \Sigma_2(u)=1+\frac12\sin^2(2u),
\end{equation}
the regular state-adding class is
\begin{equation}
 \frac{V_{\mathcal A_2}(\theta)}{k^2}
 =\frac34\sec^2u+\frac34\csc^2u
 +\frac{24}{\Delta_2(u)}-\frac{16}{\Delta_2(u)^2},
\label{eq:concise-A2-potential}
\end{equation}
with angular eigenvalues $\nu/k=1$ and $5+2m$.  By contrast,
\begin{equation}
 \frac{V_{\mathcal A_3}(\theta)}{k^2}
 =\frac34\sec^2u+\frac34\csc^2u
 +\frac{40}{\Sigma_2(u)}-\frac{48}{\Sigma_2(u)^2}+8\sec^2(2u)
\label{eq:concise-A3-potential}
\end{equation}
has an interior pole at $\theta=\pi/(4k)$ and therefore defines two separate self-adjoint interval problems.  These examples already realize the three basic possibilities needed later: ordinary/isospectral ladders, an added isolated level, and a split-domain spectrum.

\subsection{Algebraic families and parameter constraints}

For Tables 48.4--48.5 the GLS solutions are given parametrically by $x=x(s)$ and $y=y(s)$.  No inversion is required:
\begin{equation}
 y_x=\frac{y_s}{x_s},\qquad
 \theta(s)=\frac1k\arccos\sqrt{x(s)},\qquad
 V(s)=V\!\left(x(s),y(s),\frac{y_s}{x_s}\right).
\label{eq:concise-parametric-potential}
\end{equation}
In the following formula, the exponent parameters
\(\theta_0,\theta_1,\theta_t,\theta_\infty\), distinct from the angular
coordinate \(\theta\), are defined by
\(\alpha=\theta_\infty^2/2\), \(\beta=-\theta_0^2/2\),
\(\gamma=\theta_1^2/2\) and \(\delta=(1-\theta_t^2)/2\).
For the displayed square-root branch we use
\begin{equation}
 (\rho_1,\rho_2,\rho_3,\rho_4)
 =(\theta_\infty-1+\theta_t,\theta_\infty-1-\theta_t,
   \theta_0+\theta_1,\theta_0-\theta_1),
\label{eq:GLS-exponent-branch}
\end{equation}
so the exponent data determine the formal factors in Eq.~\eqref{eq:concise-Phi-PVI}.  The main patterns are collected in Table~\ref{tab:concise-algebraic-patterns}.  Table 48.4 has no real branch with $0<x<1$ in its displayed parameterization, so its entries are formal angular factorizations only. The Tables 48.5A--48.5B map $-1<s<0$ to the physical interval and therefore provide genuine starting points for a self-adjoint spectral analysis.

\begin{table}[ht]
\centering
{\footnotesize
\renewcommand{\arraystretch}{1.10}
\setlength{\tabcolsep}{4pt}
\begin{tabular}{@{}c c l@{}}
\toprule
GLS rows & $\Phi^{(1)}(N)$ up to ordering & Comment \\
\midrule
48.4: $1,2$ & $F_1^2F_0^2$ & repeated thresholds \\
48.4: $3$--$7$ & $F_0F_1F_2F_3$ & four integral factors \\
48.4: $8$ & $F_{\sqrt2}^2F_3^2$ & repeated irrational pair \\
48.4: $9,10$ & $F_3^2F_0^2$ & repeated thresholds \\
48.4: $11$ & $F_3^2F_2^2$ & repeated thresholds \\
\midrule
48.5A: $1,4$--$7$ & $F_{2/3}^2F_0^2$ & repeated thresholds \\
48.5A: $2$ & $F_1^2F_{1/3}^2$ & repeated nonintegral thresholds \\
48.5A: $3$ & $F_0F_{2/3}F_2F_{8/3}$ & four distinct factors \\
48.5A: $8$ & $F_0F_{2/3}F_{4/3}F_2$ & four distinct factors \\
48.5B: $1$--$3$ & $F_1^2F_0^2$ & repeated thresholds \\
48.5B: $4$--$6$ & $F_0F_1F_2F_3$ & four distinct integral factors \\
48.5B: $7$ & $F_1^2F_2^2$ & repeated thresholds\\
\bottomrule
\end{tabular}}
\caption{Representative formal norm-factor patterns in the algebraic GLS tables.  They determine candidate thresholds but not, by themselves, a physical spectrum.}
\label{tab:concise-algebraic-patterns}
\end{table}

The general rational-solution classification is also useful, but only one consequence is needed here.  A canonical Yuan--Li Case-II branch, with $N_{\rm Y}\in\mathbb Z$, may be written
\begin{equation}
\begin{aligned}
 \rho_1&=N_{\rm Y}+q_{\rm Y}-1,&
 \rho_2&=N_{\rm Y}-q_{\rm Y}-1,\\
 \rho_3&=N_{\rm Y}+1+2\lambda_{\rm Y}+q_{\rm Y},&
 \rho_4&=N_{\rm Y}-q_{\rm Y}-1,
\end{aligned}
\end{equation}
which implies
\begin{equation}
\rho_2=\rho_4
\qquad\Longrightarrow\qquad
 \Phi^{(1)}(N)=F_{\rho_2}(N)^2F_{\rho_1}(N)F_{\rho_3}(N).
\label{eq:concise-Yuan-collision}
\end{equation}
This repeated factor is a property of the canonical Case-II locus, not of every rational $P_{\rm VI}$ solution. The Case-I rational examples can retain four distinct thresholds \cite{yuan02,mazz01}.  Likewise the hypergeometric conditions of GLS Theorem 48.3 generally impose integral congruences between threshold families rather than an actual collision \cite{grom02}.  In all cases the formal factorization must still be supplemented by the self-adjoint and positivity analysis.

For every physical angular value $\nu$, the separated oscillator and Coulomb energies, with radial quantum number $n=0,1,2,\ldots$, are
\begin{equation}
 E^{\rm SHO}_{n;\nu}=\omega(2n+1+\nu),\qquad
 E^{\rm K}_{n;\nu}=-\frac{\kappa^2}{2(n+\nu+\frac12)^2}
=-\frac{2\kappa^2}{(2n+1+2\nu)^2}.
\label{eq:concise-energy-map}
\end{equation}

Hence an added angular singlet produces an entire additional radial tower, whereas inequivalent angular residue classes give distinct families of finite polynomial-algebra multiplets.

Table~\ref{tab:concise-GLS-spectral-classes} records which thresholds are physical, but not the mechanism behind it: why some nonclassical rational classes are strictly isospectral to P\"oschl--Teller potentials while others acquire an isolated level. Both features follow from writing these potentials as rational Darboux deformations of the P\"oschl--Teller potential, whose eigenfunctions are built from exceptional Jacobi polynomials. We develop this description in the next section, together with the corresponding spectral modules and polynomial algebras.

\section{Exceptional-Jacobi realization and spectral modules}
\label{sec:exceptional-Jacobi-Gromak}

The regular rational GLS potentials admit a sharper interpretation in terms of Darboux transformations and exceptional Jacobi polynomials \cite{gomez09,oda10,oda11,que15,ull14}.  The parameters \(\lambda_{\rm J},\mu_{\rm J}\) are the local P\"oschl--Teller/Jacobi indices, and \(\nu\) continues to denote an eigenvalue of \(N\). Put $x=\cos^2(k\theta)$ and
\begin{equation}
 V_{\rm PT}(\lambda_{\rm J},\mu_{\rm J})=k^2\left[
 \left(\lambda_{\rm J}^2-\frac14\right)\sec^2(k\theta)
 +\left(\mu_{\rm J}^2-\frac14\right)\csc^2(k\theta)\right].
\end{equation}
For a polynomial $\Xi(x)$ define
\begin{equation}
V_\Xi(\theta)=V_{\rm PT}(\lambda_{\rm J},\mu_{\rm J})
 -2\frac{d^2}{d\theta^2}\log \Xi\!\left(\cos^2(k\theta)\right).
\label{eq:concise-Darboux-form}
\end{equation}
Direct substitution shows that every nonclassical rational class from Tables 48.1--48.3 has this form.  For the interior-regular cases the eigenfunctions can be written
\begin{equation}
 \Theta_d(\theta)=
 \frac{x^{(\lambda_{\rm J}+1/2)/2}(1-x)^{(\mu_{\rm J}+1/2)/2}}{\Xi(x)}\,
 \widehat P_d(x),
\end{equation}
with positive exceptional-Jacobi weight
\begin{equation}
 w_\Xi(x)=\frac{x^{\lambda_{\rm J}}(1-x)^{\mu_{\rm J}}}{\Xi(x)^2},\qquad 0<x<1.
\end{equation}
The finite missing-degree sets and the corresponding regular spectra are summarized in Table~\ref{tab:concise-exceptional-flags}.

\begin{table}[ht]
\centering
{\small
\renewcommand{\arraystretch}{1.08}
\setlength{\tabcolsep}{4.8pt}
\begin{tabular}{@{}c c c c@{}}
\toprule
Class & $\Xi(x)$ & Polynomial degrees & Positive $\nu/k$ \\
\midrule
$\mathcal R_1$ & $x-5$ & $1,2,3,\ldots$ & $\frac{11}{2}+2m$ \\
$\mathcal R_2$ & $3x+1$ & $1,2,3,\ldots$ & $\frac72+2m$ \\
$\mathcal R_3$ & $x^2-6x-3$ & $2,3,4,\ldots$ & $\frac72+2m$ \\
$\mathcal A_2$ & $2x^2-2x+1$ & $0,3,4,\ldots$ & $1$ and $5+2m$ \\
$\mathcal A_5$ & $2x^4-4x+3$ & $0,5,6,\ldots$ & $3$ and $7+2m$ \\
$\mathcal B_2$ & $x+2$ & $1,2,3,\ldots$ & $6+2m$ \\
$\mathcal B_4$ & $3x^2+2x+1$ & $2,3,4,\ldots$ & $6+2m$\\
\bottomrule
\end{tabular}}
\caption{Regular exceptional-Jacobi sectors.  The $\mathcal R_i,\mathcal B_2,\mathcal B_4$ rows are strictly isospectral: the gaps occur in polynomial degree, not in physical energy.}
\label{tab:concise-exceptional-flags}
\end{table}

The two state-adding classes have reducible angular representations.  Their norm polynomials are
\begin{align}
 \Phi_{\mathcal A_2}(N)&=(N-k)^4\big[(N-k)^2-4k^2\big]
 \big[(N-k)^2-16k^2\big],
\label{eq:concise-Phi-A2}\\
 \Phi_{\mathcal A_5}(N)&=\big[(N-k)^2-4k^2\big]^2
 \big[(N-k)^2-16k^2\big]\big[(N-k)^2-36k^2\big].
\label{eq:concise-Phi-A5}
\end{align}
Accordingly,
\begin{equation}
\begin{aligned}
 \mathcal H_\theta^{(\mathcal A_2)}
 &=\mathbb C|\nu=k\rangle\oplus\mathcal D^+_{5k},\\
 \mathcal H_\theta^{(\mathcal A_5)}
 &=\mathbb C|\nu=3k\rangle\oplus\mathcal D^+_{7k}.
\end{aligned}
\label{eq:concise-state-adding-modules}
\end{equation}
The isolated state is annihilated by the standard $J_\pm$-based integrals but remains a physical angular eigenstate. Through Eq.~\eqref{eq:concise-energy-map} it therefore generates a complete additional oscillator or Coulomb radial tower.  The principal component participates in the usual finite-dimensional modules of the two-dimensional deformed oscillator.

\subsection{Explicit polynomial algebras for the state-adding sectors}
The difference between a state-adding system and a strictly-isospectral
exceptional system becomes particularly transparent in the complete
polynomial algebra.  Consider again the oscillator extension with
$k=\ell=p=q=1$ and
\begin{equation}
 X_+X_-
 =\Phi_{\mathcal A_i}(N)
 \bigl[H^2-\omega^2(N-1)^2\bigr],
 \qquad i=2,5.
\end{equation}
For both state-adding systems the generators satisfy
\begin{subequations}\label{eq:state-adding-common-algebra}
\begin{align}
 [A,B]&=C,\\
 [A,C]&=2\{A,B\},\\
 [B,C]&=-2B^2+\mathscr P_i(A;H),
 \qquad i\in\{2,5\}.
\end{align}
\end{subequations}
The two deformation polynomials are
\begin{align}
 \mathscr P_2(A;H)
 &=512A(A-2)\Bigl[
 20\omega^2A^2-4H^2A+40\omega^2A
 -3H^2+12\omega^2\Bigr],
 \label{eq:A2-explicit-polynomial-algebra}\\
 \mathscr P_5(A;H)
 &=512(A-6)(A-2)\Bigl[
 20\omega^2A^2-4H^2A+3H^2\Bigr].
 \label{eq:A5-explicit-polynomial-algebra}
\end{align}
These are fourth-degree deformations in the generator $A$.  Their additional
factors encode both the deleted nonphysical thresholds and the isolated
physical state.

\paragraph{The $\mathcal A_2$ module.}
The isolated angular state has $\nu=1$ and satisfies
\begin{equation}
 J_-|\nu=1\rangle=J_+|\nu=1\rangle=0,
\end{equation}
These identities follow by direct substitution of the isolated eigenfunction into the ladders, and agree with $\Phi_{\mathcal A_2}(1)=\Phi_{\mathcal A_2}(3)=0$. The polynomial zeros alone would not establish annihilation of a physical state.  The isolated state gives the
one-dimensional radial tower
\begin{equation}
 E_n^{\rm sing}=2\omega(n+1),
 \qquad n=0,1,2,\ldots.
 \label{eq:A2-singlet-tower}
\end{equation}
The principal component begins at $\nu_0=5$.  At
\begin{equation}
 E_{\mathcal N}^{\rm main}=2\omega({\mathcal N}+3),
 \qquad \nu_j=5+2j,
\end{equation}
the composite shifts act on an $({\mathcal N}+1)$-dimensional module with structure
function
\begin{equation}
 \begin{aligned}
 \varphi_{\mathcal N}^{(\mathcal A_2)}(j)
 ={}&1024\omega^2j(j+1)(j+2)^4(j+3)(j+4)\\
 &\times({\mathcal N}-j+1)({\mathcal N}+j+5).
 \end{aligned}
 \label{eq:A2-full-structure-function}
\end{equation}
For every ${\mathcal N}\geq0$, the same energy also occurs in the singlet tower at radial
quantum number $n={\mathcal N}+2$.  Hence the full eigenspace decomposes as
\begin{equation}
 \mathcal H_{E_{\mathcal N}}^{(\mathcal A_2)}
 \cong\mathbf 1\oplus\mathcal V_{{\mathcal N}+1}^{(\mathcal A_2)}.
 \label{eq:A2-energy-module-decomposition}
\end{equation}
where the polynomial integrals vanish on $\mathbf1$ and act irreducibly on
$\mathcal V_{{\mathcal N}+1}^{(\mathcal A_2)}$.  The two lowest energies $2\omega$ and
$4\omega$ contain only the singlet module.

\paragraph{The $\mathcal A_5$ module.}
Here the isolated state has $\nu=3$. Direct substitution again shows that both angular ladders annihilate it, consistently with $\Phi_{\mathcal A_5}(3)=\Phi_{\mathcal A_5}(5)=0$.  Its radial tower is
\begin{equation}
 E_n^{\rm sing}=2\omega(n+2),
 \qquad n=0,1,2,\ldots.
 \label{eq:A5-singlet-tower}
\end{equation}
The principal ladder starts at $\nu_0=7$.  The energy and structure function are
\begin{equation}
 E_{\mathcal N}^{\rm main}=2\omega({\mathcal N}+4),
 \qquad \nu_j=7+2j,
\end{equation}
\begin{equation}
 \begin{aligned}
 \varphi_{\mathcal N}^{(\mathcal A_5)}(j)
 ={}&1024\omega^2j(j+1)(j+2)^2(j+4)^2(j+5)(j+6)\\
 &\times({\mathcal N}-j+1)({\mathcal N}+j+7).
 \end{aligned}
 \label{eq:A5-full-structure-function}
\end{equation}
Thus, for ${\mathcal N}\geq0$,
\begin{equation}
 \mathcal H_{E_{\mathcal N}}^{(\mathcal A_5)}
  \cong\mathbf 1\oplus\mathcal V_{{\mathcal N}+1}^{(\mathcal A_5)}
 \label{eq:A5-energy-module-decomposition}
\end{equation}
with the singlet at radial quantum number $n={\mathcal N}+2$.  The levels $4\omega$ and
$6\omega$ lie below the principal tower and consist only of one-dimensional
modules.  These decompositions are the algebraic signature of state addition:
the added angular state is physical but invisible to the standard composite
integrals.

The singular class $\mathcal A_3$ is qualitatively different.  Its Darboux denominator contains $2x-1$, and near
\begin{equation}
 \theta_*=\frac{\pi}{4k}
\end{equation}
one has
\begin{equation}
 V_{\mathcal A_3}(\theta)=\frac{2}{(\theta-\theta_*)^2}+O(1).
\label{eq:concise-A3-barrier}
\end{equation}
The interval therefore splits into two self-adjoint problems.  The formal polynomial Heisenberg relation
\begin{equation}
 [N,J_\pm]=\pm2kJ_\pm,\qquad
 J_+J_-=\Phi(N),\qquad J_-J_+=\Phi(N+2k)
\end{equation}
continues to hold as a formal operator identity, but a physical ladder representation exists only if the operators preserve the chosen self-adjoint domain. This is the sharpest illustration in the catalogue of the distinction between formal Painlev\'e factorization and quantum spectrum.

Taken together, the rational examples realize four spectral types: ordinary Jacobi ladders, strictly isospectral exceptional-Jacobi ladders, state-adding systems with an isolated angular level, and a split-domain singular problem. In all of them the ladder and polynomial algebras of Sections~2--8 hold as formal identities, while the physical representation is selected by the domain and the boundary conditions. We summarize these results in the final section.

\section{Conclusions and outlook}
\label{sec:conclusion}

We have constructed generalized ladder operators for quantum Hamiltonians with a quadratic spectral variable and applied them to the Painlev\'e VI angular Hamiltonian.  This is a new result, and it completes previous works where Painlev\'e I, II, IV and V were related to operator algebras \cite{saj19,mar19a}. The resulting fourth-order ladders close through an eighth-degree norm polynomial and, after combination with the radial shifts of the oscillator, Coulomb and free Bessel systems, generate higher-order integrals and deformed Racah-type polynomial algebras for two-dimensional superintegrable models. This allows us to obtain families of Hamiltonians that share similarities with the Tremblay--Turbiner--Winternitz (TTW) \cite{ttw09,ttw10} and Post--Winternitz (PW) \cite{pw10} systems, extending these systems, which involve only trigonometric functions, to families involving higher transcendental functions. This framework recovers, as limits, known Hamiltonians from the classifications of superintegrable systems separating in polar coordinates with a third-order \cite{tre10} or fourth-order \cite{esc17,esc18a} integral and, as very special limits of the Painlev\'e VI parameters, models such as the TTW and PW systems, as well as a range of their deformations \cite{post12a,rita,ritb}. As Painlev\'e VI admits classical solutions expressed through hypergeometric functions \cite{iwasaki91,watanabe98}, is closely related to the Heun equation \cite{slavyanov96,derezinski21} and possesses families of rational and algebraic solutions \cite{mazz01,yuan02,boalch06,lisovyy14}, our framework provides in a unified way a large class of superintegrable systems that would otherwise be obtained in an ad hoc way. Previous constructions of superintegrable systems from operator algebras were only achieved for Painlev\'e I, II, IV and V \cite{saj19} and for rational Painlev\'e VI potentials \cite{rita,ritb}. Here the construction of superintegrable systems for a generic Painlev\'e VI transcendent based on an underlying operator algebra is new, and fills a gap in the literature. It also points out how quantum mechanics, the spectral problem, the related operator algebras and deformed Racah algebras are intimately intertwined with Painlev\'e transcendents.

For the oscillator realizations we obtained the Casimir relation and recovered the finite-dimensional spectra directly from the deformed-oscillator structure function \cite{defosc1,defosc2}. The Gromak--Laine--Shimomura (GLS) special solutions \cite{grom02} make the representation content of this construction explicit.  After the parameter conversion in Eq.~\eqref{eq:concise-parameter-dictionary}, the rational examples collapse into a small number of spectral types: ordinary Jacobi ladders, strictly-isospectral exceptional-Jacobi ladders, state-adding systems with an isolated angular level, and a split-domain singular problem.  The two branches of $\rho_1+\rho_2=-2\pm\sqrt{8\alpha}$ can correspond to genuinely different Hamiltonians, while sign changes of the other three square-root combinations in \eqref{eq:concise-alpha-branches} leave the quantum potential unchanged.  The state-adding examples also show that a single isolated angular level becomes an entire additional radial spectral tower in the full two-dimensional system.  Their explicit fourth-degree polynomial algebras decompose each overlapping energy eigenspace into a one-dimensional singlet and an irreducible principal multiplet. This points out the possibility of designing a wide range of interactions by selecting special values of the four parameters of the Painlev\'e VI transcendent, and of obtaining much richer spectra than those of the hydrogen atom, the radial oscillator and many of their deformations.

\section*{Data availability}
No new data were created or analysed in this study. 

\section*{Acknowledgements}
IM was supported by the Australian Research Council Future Fellowship FT180100099. 

\printbibliography[title={References}]
\end{document}